\documentclass[prb,showpacs,twocolumn,amsmath,amssymb,floatfix,longbibliography]{revtex4-1}
\usepackage{graphicx}
\usepackage{bm}
\usepackage{bbm}
\usepackage{color}
\usepackage{footnote}
\usepackage{epstopdf}
\usepackage{hyperref}
\usepackage{appendix}
\usepackage[applemac]{inputenc}
\newcommand{\new}[1]{#1}

\begin{document}
\title{SNS junctions in nanowires with spin-orbit coupling: role of confinement and helicity on the sub-gap spectrum}
\author{Jorge Cayao$^1$, Elsa Prada$^2$, Pablo San-Jose$^1$ and Ramón Aguado$^1$}
\affiliation{$^1$Instituto de Ciencia de Materiales de Madrid (ICMM-CSIC), Cantoblanco, 28049 Madrid, Spain\\$^2$Universidad Autónoma de Madrid, Cantoblanco, 28049 Madrid, Spain}
\date{\today} 
\begin{abstract}
We study normal transport and the sub-gap spectrum of superconductor-normal-superconductor (SNS) junctions made of semiconducting nanowires with strong Rashba spin-orbit coupling. We focus, in particular, on the role of confinement effects in long ballistic junctions. In the normal regime, scattering at the two contacts gives rise to two distinct features in conductance, Fabry-Perot resonances and Fano dips. The latter arise in the presence of a strong Zeeman field $B$ that removes a spin sector in the leads (\emph{helical} leads), but not in the central region. Conversely, a helical central region between non-helical leads exhibits helical gaps of half-quantum conductance, with superimposed helical Fabry-Perot oscillations. These normal features translate into distinct subgap states when the leads become superconducting. In particular, Fabry-Perot resonances within the helical gap become parity-protected zero-energy states (parity crossings), well below the critical field $B_c$ at which the superconducting leads become topological.  As a function of Zeeman field or Fermi energy, these zero-modes oscillate around zero energy, forming characteristic loops, which evolve continuously into Majorana bound states as $B$ exceeds $B_c$. The relation with the physics of parity crossings of Yu-Shiba-Rusinov bound states is discussed.
%
%
%\redout{In the normal regime and} In the presence of a finite Zeeman field, we find two \new{distinct} kinds of resonances \new{in normal conductance}, Fabry-Perot resonances and Fano dips. 
%\redout{with distinct properties in the helical regime}.
%When the contacts are superconducting, these \redout{helical} resonant features give rise to novel sub-gap states \redout{with non-trivial behavior}. 
%Perhaps most importantly, the sub-gap spectrum shows multiple loops around zero energy with respect to the Fermi energy and the Zeeman field in regions where the N side is helical but well below the critical Zeeman field $B_c$ at which the S regions become topological. These oscillating near-zero subgap states in the trivial regime are smoothly connected to Majorana bound states when the Zeeman field is increased beyond $B_c$.
%The relation with the physics of parity crossings of Yu-Shiba-Rusinov bound states is discussed.
%as the system undergoes into the topological superconducting phase hosting Majorana fermions driven by a Zeeman field. 
%As we know, the interplay of all of these ingredients allow us to study exotic physics as it is the physics of p-wave superconductors where Majorana fermions, in condensed matter systems, were predicted to appear. 
%We explicitly show how the Andreev levels formed at the junction evolve into Majorana bound states and study the role of the finite-length effects.
\end{abstract}
\maketitle
\section{Introduction}
Majorana fermions, particles that are their own antiparticles, have been the subject of intense research over the past decades in the context of particle physics and cosmology\cite{majorana, Avignone:RMP08}. During the last few years, this interest extended to the condensed matter arena where Majorana fermions are intensely studied nowadays \cite{wilcek,Elliott-Franz}. This state of affairs has been driven by the key observation that emergent quasiparticles in superconductors can be described as Majorana fermions \cite{Alicea:RPP12,Leijnse:SSAT12,Beenakker:ARCMP13,Stanescu:JOPCM13,Elliott:14}. This, together with the recent advances in the field of topological materials \cite{Hasan:RMP10,Qi:RMP11}, has  spurred an intense search for condensed matter realizations of Majorana fermions. Most of these realizations focus on zero-energy modes inside the gap of topological superconductors. These zero modes are Majoranas from the point of view of particle-antiparticle conjugation, but they do not obey fermionic exchange statistics  \footnote{In fact they obey non-Abelian exchange statistics which might have potential applications in fault-tolerant quantum computation. See, Chetan Nayak, Steven H. Simon, Ady Stern, Michael Freedman, and Sankar Das Sarma, ``Non-abelian anyons and topological quantum computation'', Rev. Mod. Phys. 80, 1083–1159 (2008)}. Thus, instead of Majorana fermions, they are now more precisely referred to as Majorana bound states (MBSs) or Majorana zero modes. \footnote{Recently, it has been argued that Bogoliubov quasiparticles in conventional superconductors are true Majorana fermions. See, C, Chamon, R. Jackiw, Y. Nishida, S.-Y. Pi and L. Santos,  ``Quantizing Majorana fermions in a superconductor'', Phys. Rev. B {\bf 81}, 224515, (2010). Their Majorana fermion nature can be revealed by annihilation processes, see C. Beenakker, ``Annihilation of Colliding Bogoliubov Quasiparticles Reveals their Majorana Nature'', Phys. Rev. Lett. {\bf 112}, 070604 (2014)}.

%but also exhibit amazing properties to be potentially used in fault-tolerant quantum computation\cite{kitaev, RevModPhys.80.1083}, but also because they represent the most simple realization of a new family of exotic states, that could open new horizons in our understanding of fundamental physics.
%Although originally predicted to appear in particle physics \cite{majorana, RevModPhys.80.481}, it seems that their condensed matter counterparts, typically zero modes with MF character, are closer to be detected \cite{wilcek, stern, Physics.3.24, Service08042011, Physics.4.67}. 

Early proposals suggested that MBSs can emerge in exotic superconductors, such as p-wave, since they realize topological phases that support edge excitations with Majorana fermion character \cite{PhysRevB.44.9667, volovik, PhysRevB.61.9690, PhysRevB.61.10267, kitaev, PhysRevB.73.220502}. 
Even though p-wave pairing is not robust against disorder and thus scarce in nature, one can engineer systems to mimic such non trivial superconductivity. These are based on the proximity effect between a conventional s-wave superconductor and a topological insulator \cite{Fu:PRL08}, or a semiconductor nanowire  (NW) with strong spin-orbit (SO) coupling \cite{Sato:PRL09,Sau:PRL10,Alicea:PRB10,Lutchyn:PRL10,Oreg:PRL10}. For the latter case it has been shown \cite{Lutchyn:PRL10,Oreg:PRL10} that if an external Zeeman field $B$, orthogonal to the SO axis, exceeds a critical value $B_{c}\equiv\sqrt{\mu^2+\Delta^2}$, where $\mu$ is the Fermi energy and  $\Delta$ the induced s-wave pairing, zero energy MBSs emerge at the nanowire ends signaling a topologically non-trivial phase.
%Although it is cleat from the theoretical point of view that Majorana modes do exist in topological superconductors under the right conditions (such as the above NW proposal), 

Unfortunately, the outcome of the simplest detection protocol for MBSs in NW devices \cite{Sengupta:PRB01,Bolech:PRL07,Law:PRL09}, detection of subgap zero modes through zero-bias anomalies in transport  \cite{Mourik:S12,Deng:NL12,Das:NP12,Finck:PRL13,Churchill:PRB13,Lee:13}, can be obscured, or even mimicked, by other effects \cite{Lee:PRL12,Pientka:PRL12,Bagrets:PRL12,Liu:PRL12,Sau:13,Prada:PRB12,Kells:PRB12,Zitkoetal}.  
%The first  zero-bias anomaly (ZBA) measurements from Delft  \cite{frolov,xu,Das:NP12,Finck:PRL13,Churchill:PRB13,Lee:13} in the differential conductance $dI/dV$ are consistent with the existence of such zero energy MBSs \cite{Sengupta:PRB01,Bolech:PRL07,Law:PRL09,Prada:PRB12} and further experiments from the same group show the expected robustness versus external parameters expected from MBSs \footnote{Leo Kouwenhoven, private communication}. However, subsequent experiments show a somewhat less robust phenomenology \cite{xu,Das:NP12,Finck:PRL13,Churchill:PRB13}. Importantly, essentially the same same phenomenology may arise owing to Kondo physics\cite{Lee:PRL12, Finck:PRL13,Churchill:PRB13,Zitkoetal}, disorder\cite{Pientka:PRL12,Bagrets:PRL12,Liu:PRL12,Sau:13},  smooth confinement \cite{Prada:PRB12,Kells:PRB12}, parity crossings of Andreev levels \cite{Lee:13,Zitkoetal}, etc.
 As a result, there is no clear consensus yet on whether MBSs have been observed or not in NWs \footnote{ Quite recently, further evidence of zero-bias anomalies related to Majoranas have been reported in a different setup consisting of a ferromagnetic atomic chain on top of a superconducting substrate. S. Nadj-Perge, {\it et al}, ``Observation of Majorana fermions in ferromagnetic atomic chains on a superconductor'', Science, {\bf 346}, 602 (2014).}.

Thus, the time seems right to move beyond zero-bias anomaly experiments and study more complex geometries such as Superconductor-Normal-Superconductor (SNS) junctions \cite{pablo,PhysRevLett.112.137001,Flensberg-Xu}. This geometry has a number of advantages including the possibility of studying supercurrents \cite{Deng:NL12,Doh:S05,Nishio:N11,Nilsson:NL12,Gunel:JOAP12}, or direct spectroscopy of Andreev bound states (ABS) \cite{PhysRevLett.110.217005, nphys1811, PhysRevLett.104.076805, PhysRevB.88.045101,PhysRevB.89.045422,PhysRevX.3.041034,Levy,Lee:13,PhysRevLett.85.170}.  As we shall discuss, this latter technique can be used, in principle, to directly monitor the detailed evolution from the trivial to the nontrivial regime. Previous papers have mostly focused on short 
junctions\cite{pablo,Bena1,gilbertini2012,Bena2,stefano2014}, while detailed studies of ABS in other relevant geometries, including long and intermediate-length junctions, remain largely unaddressed. In particular, the role of Fabry-Perot resonances occurring in normal transport as the middle NW finite-lenght section of the junction is depleted has never been studied to the best of our knowledge. In this work we fill this void and present detailed calculations of the normal conductance and Andreev spectra in such geometries. %Our results are relevant for nanowire-based SNS topological junctions containing confined levels. 
We emphasize here that all nanowire experiments should ideally belong to the category studied here, as confinement effects should be present when a ballistic quasi-one dimensional conductor is contacted between leads, especially when the normal part of the NW (in our geometry, the region of length $L_{nw}$ not directly in contact with leads) is gated. This electrical gating naturally creates quantum wells (or barriers) with their associated confined quantum levels in the middle region of the NW.

\new{In the first half of this work, we discuss normal transport across a finite length ballistic NW. We show how bandstructure details in the presence of strong Rashba SO coupling and Zeeman fields may dominate transport, and give rise to distinct features associated to helical phases (defined by singly-degenerate subbands at the Fermi level with spin locked to momentum) known as helical gaps (Fig.\,\ref{fig1}). Likewise, finite contact resistance induces confinement resonances in conductance as quasibound states develop in the NW}. In the simplest case of non-interacting electrons \footnote{Coulomb blockade effects will be discussed elsewhere.}, \new{we find that confinement generates two types of resonances: Fabry-Perot resonances and helical Fano dips.  Fabry-Perot resonances for a spinful mode \cite{Kretinin:NL10} will give conductance oscillations with a ceiling of $2e^2/h$, unless the central NW is depleted into its helical regime, in which case one may observe \emph{helical} Fabry-Perot resonances with a half-quantum $e^2/h$ ceiling.  For long enough junctions, many helical Fabry-Perot resonances may occur. We discuss that, while confinement effects may mask the helical gap, the characteristic reentrance of helical Fabry-Perot resonances with Zeeman field or gate voltage contains valuable information about non-trivial helical transport through the NW. The second kind of resonances are sharp Fano dips when the central section of the NW is gated to form a quantum well (non-helical) and the NW sections below the contacts (the leads) are helical.} Therefore, both types of resonant features in normal transport may signal the helical regime in different sections (central or below the contacts) of the NW.  In the presence of superconducting leads, the two lead to distinct effects.
%As opposed to the Fabry-Perot case, they appear only when the NW sections \emph{below the contacts} are helical.

%The presence of Fabry-Perot resonances \cite{Kretinin:NL10} in particular may mask the visibility of helical gaps \cite{PhysRevB.90.235415}. 
%can greatly affect electronic transport. For example, the expected helical gap (Fig. 1) in the conductance plateaus of quantum point contacts made of NWs with strong Rashba SO coupling \cite{helical} can be masked by Fabry-Perot resonances \cite{PhysRevB.90.235415}. 
%Nevertheless, as we discuss here, these resonances still contain valuable information about non-trivial helical transport in the NW, and may exhibit reentrant behaviour of the normal conductance, \new{revealing the presence of a helical gap despite the confinement}.  %These dips are reminiscent of the so-called Fano-Rashba resonances in NWs with spatially varying couplings \cite{PhysRevB.74.153313}. 

\new{In the second half of this work we consider the connection of this phenomenology to transport in the superconducting regime.} Each helical Fano dip in the normal phase translates, in an SNS geometry, into a single subgap state that crosses zero energy as a function of external parameters (Fermi energy or Zeeman field). Such a crossing is often known as a \emph{parity crossing}, since it is protected by conservation of number parity in the junction. As we discuss, these parity crossings are made possible by the nontrivial topology in the underlying effective p-wave superconductor for $B>B_c$. Similar bound states originated from nonmagnetic impurities in topological superconductors and superfluids have been recently discussed  in Refs.  \onlinecite{Sau-Demler,Huetal} and can be considered the p-wave counterparts of Yu-Shiba-Rusinov bound states \cite{subgap1,subgap2,subgap3,subgap4,subgap5} in standard s-wave superconductors with magnetic impurities.
%The spin-resolved Fabry-Perot resonances are a distinct signature of the (finite length) normal part entering into the helical regime. 
A more direct analogy with standard Yu-Shiba-Rusinov magnetic bound states actually applies in the non-topological phase $B<B_c$. In this situation, helical Fabry-Perot resonances in normal conductance translate, in the superconducting regime, into loops around zero energy in the ABS spectrum as a function of external parameters. For long junctions, many of these loops are visible, each separated by a parity crossing at zero energy. As a result, the $B<B_c$ subgap spectrum contains near-zero energy subgap states that oscillate as a function of Fermi energy or Zeeman field when the N region of the junction is helical. Interestingly, we find that these oscillating near-zero subgap states in the trivial regime are smoothly connected to MBS when Zeeman is increased beyond $B_c$.
%\footnote{See also Hui Hu, Han Pu, Yan Chen and Xia-Ji Liu, Phys. Rev. Lett. {\bf 110}, 020401 (2013) and J. D. Sau and E. Demler, Phys. rev. B {\bf 88} 205402 (2013)}.  

%The critical current reflects this nontrivial resonant behavior in the junction. Spin-resolved Fabry-Perot resonances give rise to $\pi$-junction behavior in agreement with our interpretation in terms of magnetic Shiba states. Helical Fano dips, on the other hand, give rise to...

This paper is organized as follows. In section \,\ref{sec1} we describe the Hamiltonian model employed in our work. 
%For completeness, we discuss 
%how the interplay of Zeeman magnetic field and s-wave superconductivity in NWs with Rashba SOC mimic the physics of p-wave pairing, thus allowing us to investigate the existence of MBSs. We point out that in the non-topological phase, the problem can be interpreted in terms of two independent p-wave superconductors, originated from the Rashba helical bands, and weakly coupled by an interband pairing term, while the situation switches to the presence of one p-wave superconductor in the topological phase.
Section \ref{normCon} focuses on the normal conductance and how the two types of resonances, helical Fabry-Perot and helical Fano dips, appear in the system. The rest of the paper is devoted to analysing the consequences of these resonant levels in the sub-gap spectrum in the superconducting regime.
% In section \ref{ElevelsNS} we study in detail the subgap spectrum in a I-N-S configuration, similar to the ones used to study ZBA formation. Different lengths of the normal section are shown to give very different phenomenology in the subgap spectrum, including oscillating near-zero energy subgap states. We also discuss how the helical dips of the normal phase lead to parity crossings when the NW is an effective p-wave topological superconductor.
After a brief discussion on how the SNS junction is modeled, as well as a discussion about the relevant length scales of the problem, section \ref{ElevelsSNS} presents a systematic study of the subgap spectrum of SNS junctions, \new{including its dependence on the superconducting phase difference across the junction, $\varphi$.} 
%In this case, the superconducting phase difference across the junction, $\varphi$, provides an extra control knob to study the subgpap spectrum. 
We discuss in detail how the presence of confined levels within the central region affect the ABS and lead to parity crossings in the topological phase.  The dependence of the ABS on phase difference, Fermi energy of the normal region and Zeeman field is discussed for both short and long junctions in subsections \ref{short} and \ref{long}, respectively. 
Our conclusions are presented in Section \ref{concl}. In appendix \ref{appA0} we describe in detail how we model SNS junctions by using a tight-binding version of the model resented in Section \ref{sec1}.  Appendix \ref{tightc} discusses an effective model that fully explains the phenomenology behind helical Fano resonances. 
\section{Nanowire Model}
\label{sec1}
%\subsection{SNS junctions}
%\subsection{Rashba nanowire}
\new{We present the model for a nanowire with Rashba SOC and in the presence of an external Zeeman. We restrict ourselves to the strictly one dimensional (single-mode) case for simplicity. Generalisations to multimode nanowires are relatively straightforward. The model Hamiltonian reads}
\begin{equation}
\label{Leq1}
H_{0}\,=\,\frac{p^{2}}{2m^{*}}\,-\,\mu\,-\,\frac{\alpha_{R}}{\hbar}\,\sigma_{y}\,p\,+\,B\,\sigma_{x}\,,
\end{equation}
where $p$ is the momentum operator, $m^{*}$ is the effective electron mass, $\alpha_{R}$ the Rashba SOC strength, $\mu$ the Fermi energy and $\sigma_{i}$ the spin Pauli matrices. An external magnetic field $\mathcal{B}$ along the wire produces a Zeeman splitting $B=g\mu_{B}\mathcal{B}/2$, where $\mu_{B}$ is the Bohr magneton and $g$ the wire $g$-factor.  The Rashba coupling defines a typical length, the spin-orbit length $l_{SO}\equiv\hbar/\sqrt{2m^*E_{SO}}$, with the spin-orbit energy defined as $E_{SO}=\frac{1}{2}\alpha_{R}^2m^{*}/\hbar^2$. For typical InSb values
$m^{*}=0.015\,m_{e}$,  with $m_{e}$ the electron mass and $\alpha_{R}=0.2$ eV \AA, the spin-orbit energy is $E_{SO}\approx 50 \mathrm{\mu eV}$ which gives SO lengths of the order of $l_{SO}\approx 200$nm. 

Note that the Rashba and Zeeman fields in Eq.\,(\ref{Leq1}) are perpendicular. As a result, the two spinful bands (shifted by SO) become mixed by the Zeeman term and the zero-field crossing point at zero momentum becomes an anticrossing of size $2B$. When the chemical potential lies within this anti crossing gap, the system has two Fermi points, as opposed to four Fermi points for $\mu$ above or below this gap. This window is a \emph{helical} gap, since the two fermi points correspond to counter propagating states with different spins (the spin projection is locked to momentum) \cite{helical}, see inset in Fig.\,\ref{fig1}. 

%In our calculations we consider, unless indicated otherwise, that the left and right leads L(R) have the same chemical potentials $\mu_{L(R)}=\mu$ and the same length $L_{L(R)}=L_{S}$, and $\tau=1$

\section{The normal conductance}
\label{normCon}
  \begin{figure}[!ht]
%\begin{minipage}[t]{\linewidth}
\centering
\includegraphics[width=.5\textwidth]{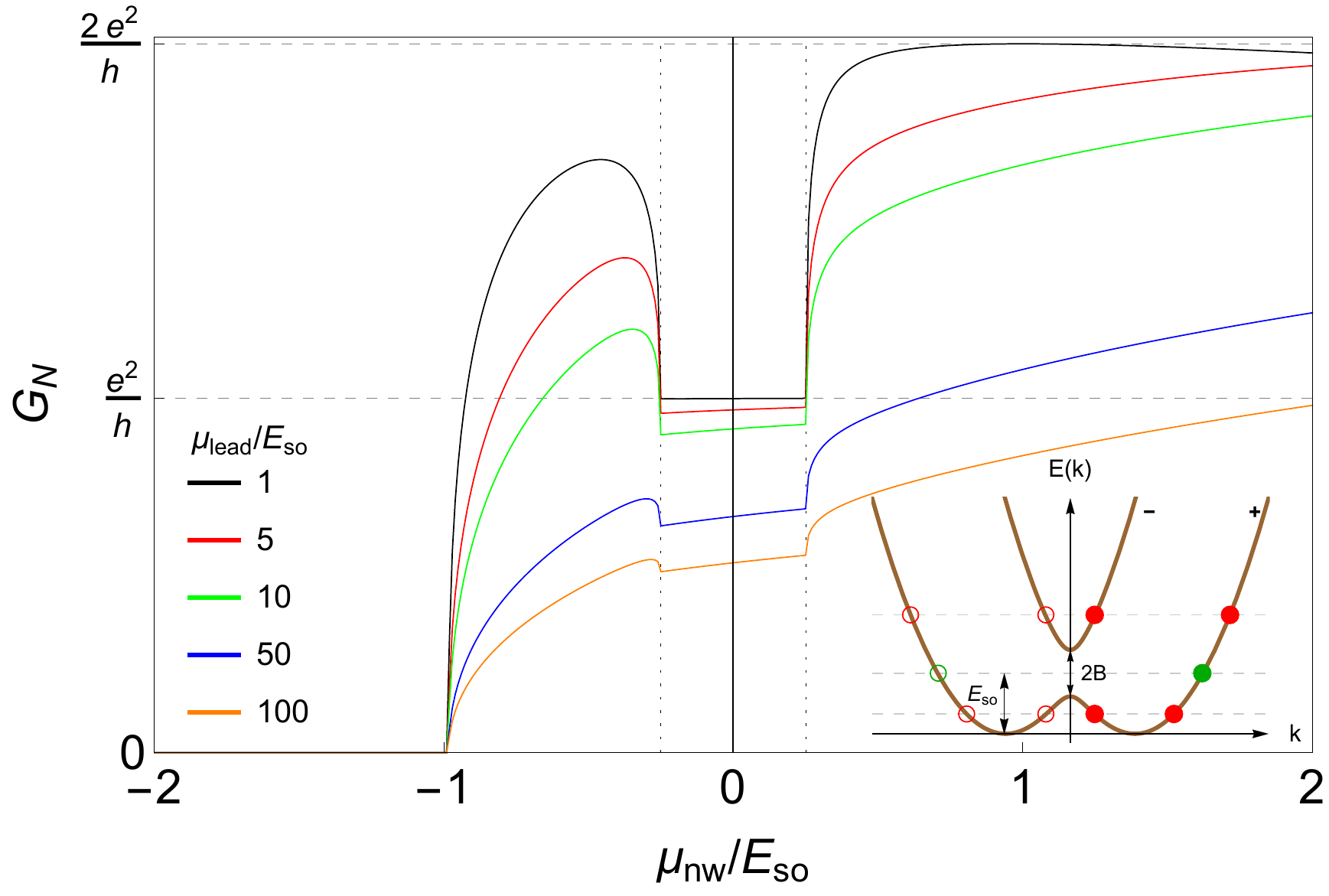} 
\caption{(Color online) Normal conductance $G_{N}$ as a function of the Fermi energy \new{$\mu_{\mathrm{nw}}$ in the left lead} for a semi-infinite NW-N junction. Parameters: $\alpha_{R}=20$\,meV\,nm (which corresponds to $E_{SO}=0.05$\,meV) and $B=0.0125$\,meV. Different curves show how $G_{N}(\mu_{\mathrm{nw}})$ evolves for increasing Fermi energy \new{$\mu_{\mathrm{lead}}$ in the right lead}. The inset shows the dispersion relation for a Rashba NW in the presence of a transverse $B$ field. Within the gap there is only one right mover per energy (green filled circle), while outside the gap there are two (red filled circles). This gives rise to the reentrant behavior of conductance, from $\sim 2e^2/h$ to $e^2/h$ and back to $2e^2/h$, as a function of Fermi energy in the main panel. The spin of the counter propagating states (open circles) is opposite to the propagating ones (filled circles), hence the name helical.}
\label{fig1}
%\end{minipage}
\end{figure}
Before discussing the sub-gap \new{Andreev spectrum of a NW} coupled to superconducting leads, we characterize the normal regime in the presence of a Zeeman field. \new{We are interested in particular in the} normal conductance $G_{N}$ as the Fermi energy ($\mu_{\mathrm{nw}}$) in the middle section of the NW (length $L_{nw}$) varies with respect to the one in the left and right leads $\mu_{\mathrm{leads}}$. Such situation models a NW contacted between normal electrodes and with a Fermi energy tuned by a central gate, see e. g. Ref. \onlinecite{Churchill:PRB13}. For simplicity in the discussion, we model the gate-induced electrostatic potential with an abrupt profile (the role of smooth gate potentials has been recently discussed in Ref. \onlinecite{PhysRevB.90.235415}).

%Although here we study a normal system (without superconductivity), we will see that the levels bound by increasing $\mu_{N}$ reveals an important information when one investigate junctions formed by linking superconductors.
%It was shown that at $B\equiv\mu$ the left and right regions experiment a transition into a helical phase\cite{helical}.
For computations purposes we discretize Eq.\,(\ref{Leq1}) into a tight-binding lattice.
The momentum operator introduces hopping elements $v$ between nearest-neighbor sites. 
The transparency of the left and right contacts is parameterised by a factor $\tau\in[0,1]$, introduced in the hopping matrix $v_{0}=\tau v$ across the two interfaces, see Appendix \ref{appA0}.
%This coupling is parametrized by a hopping matrix $v_{0}=\tau v$ between the sites that define the interfaces of the SNS junction, where $\tau\in[0,1]$, see Appendix \ref{appA0} for more details.
$G_{N}$ is calculated by means of the Greens function technique\cite{Caroli,Haug:07}, 
\begin{equation}
\begin{split}
 G_{N}&=4\frac{e^{2}}{h}\,{\rm Tr}[\Gamma_{L}\,G^{r}\,\Gamma_{R}\,G^{a}]\\
   \end{split}
 \end{equation}
 where $G^{r}=g_{0}^{r}+g_{0}^{r}\,\Sigma^{r}\,G^{r}=(G^{a})^{\dagger}$ is the full retarded Green's function. The bare Green's function of the normal region without the presence of the leads is $g_{0}^{r}=[\omega-h_{0}+i 0^{+}]^{-1}$. The hamiltonian $h_{0}$ corresponds to $H_{0}$ in Eq.\,\ref{Leq1} with $\mu=\mu_{\mathrm{nw}}$. The leads are taken into account through the self-energies $\Sigma_{L(R)}^{r}=v\,g^{r}_{L(R)}v^{\dagger}$, where $g_{L(R)}^{r}=[\omega-h_{L(R)}+i 0^{+}]^{-1}$ stands for the left/right lead's propagator, when decoupled from the system. In this case, $h_{L(R)}$ corresponds to $H_{0}$ in Eq.\,(\ref{Leq1}) with $\mu=\mu_{\mathrm{leads}}$.  Finally, $\Gamma_{L(R)}=\frac{\Sigma_{L/R}^{r}-\Sigma_{L/R}^{a}}{2i}$. In practice, $G_{N}$ is computed recursively with the boundary conditions imposed by the leads. 
 \begin{figure}[!ht]
%\begin{minipage}[t]{\linewidth}
\centering
\includegraphics[width=.5\textwidth]{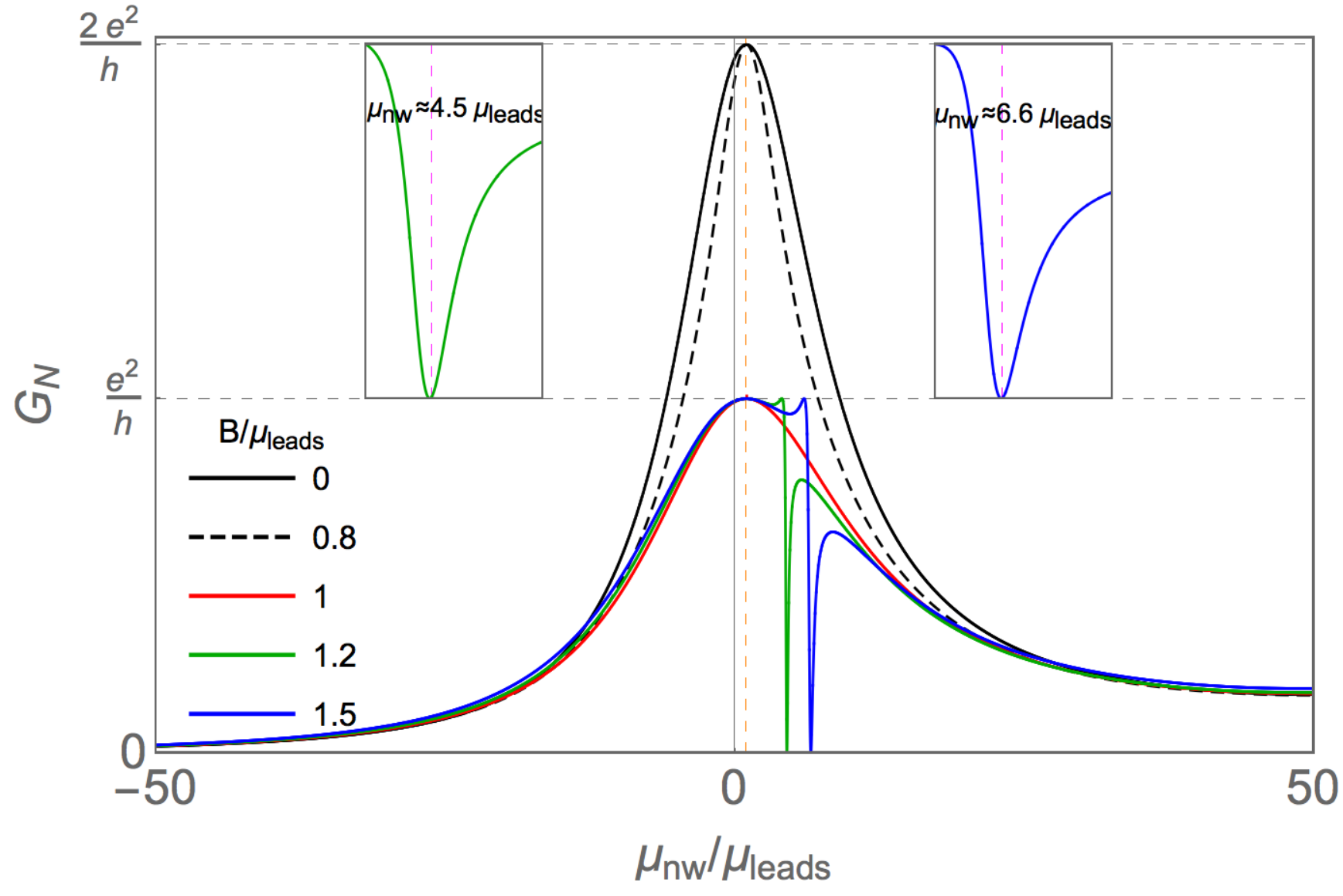} 
\caption{(Color online) Normal conductance $G_{N}$ as a function of the Fermi energy $\mu_{\mathrm{nw}}$ for a short \new{N-NW-N junction, $L_{\mathrm{nw}}=20$nm} (rest of parameters $E_{SO}=0.05$\,meV and $\mu_{\mathrm{leads}}=10E_{SO}$). Different curves show how $G_{N}(\mu_{\mathrm{nw}})$ evolve with the Zeeman field $B$.
% solid black curve $B=0$, dashed black $B=0.8\mu$, red curve at the transition $B=\mu$, in the helical phase the green curve in the left panel $B=1.2\mu$ and the blue one $B=1.5\mu$. The vertical dashed orange line shows the conductance maximum, when $\mu_{N}=\mu$. 
The insets show a blow-up of $G_{N}(\mu_{\mathrm{nw}})$ around the Fano dip for two different $B$.}
\label{fig2}
%\end{minipage}
\end{figure}
 \begin{figure}[!ht]
%\begin{minipage}[t]{\linewidth}
\centering
\includegraphics[width=.5\textwidth,height=.5\textwidth]{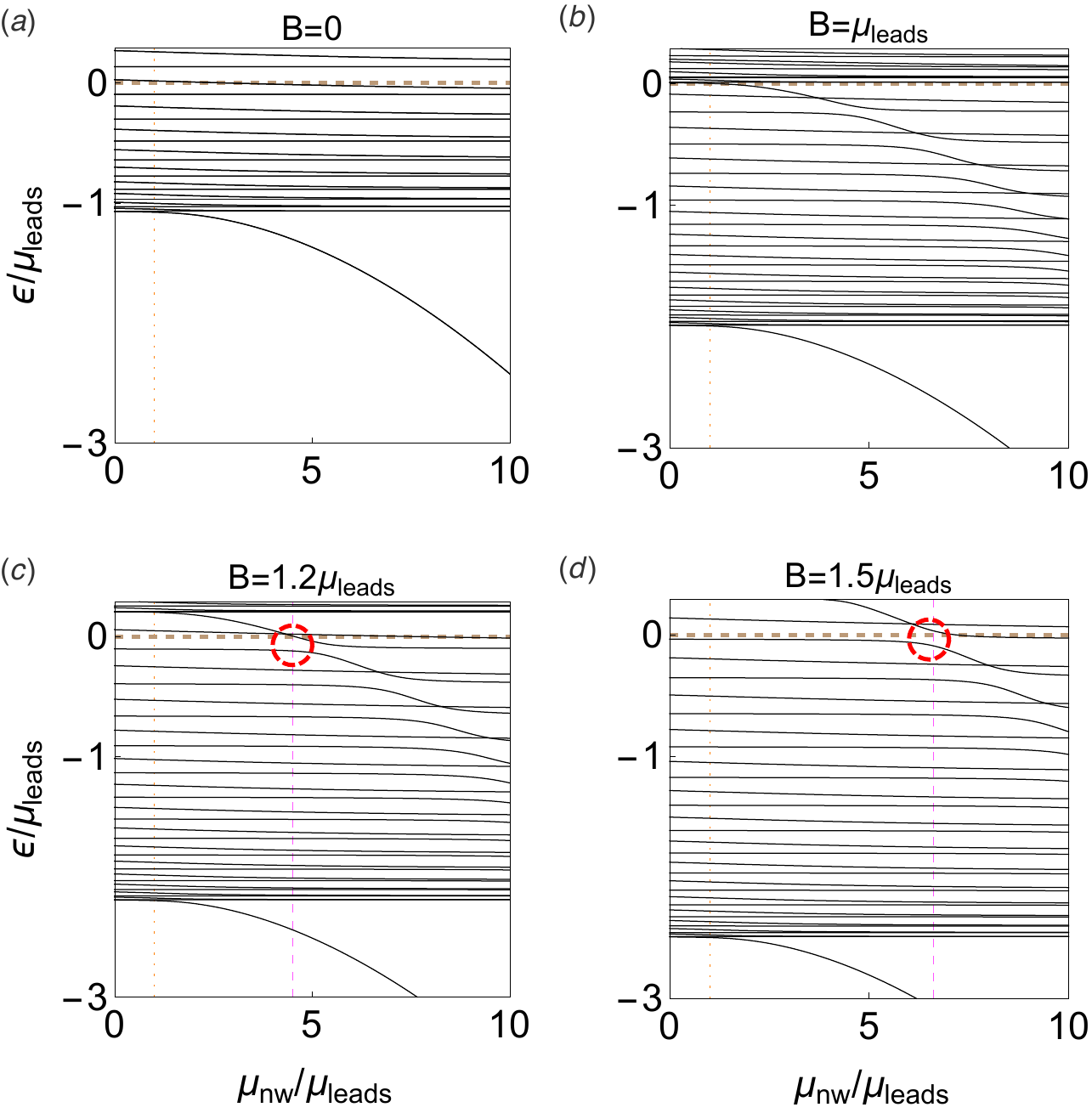} 
\caption{(Color online) Energy levels as a function of the Fermi energy $\mu_{\mathrm{nw}}$ for the same system as in Fig.\,\ref{fig2}. Different panels show how the levels evolves with the Zeeman field $B$. The red dashed circle shows the \new{value of $\mu_{\mathrm{nw}}$ for which one of the projections of the Zeeman-split bound state resonates with carriers at the Fermi level (horizontal dashed line), leading to a Fano resonance in conductance}.}
\label{fig3}
%\end{minipage}
\end{figure}
 \begin{figure}[!ht]
%\begin{minipage}[t]{\linewidth}
\centering
\includegraphics[width=.5\textwidth]{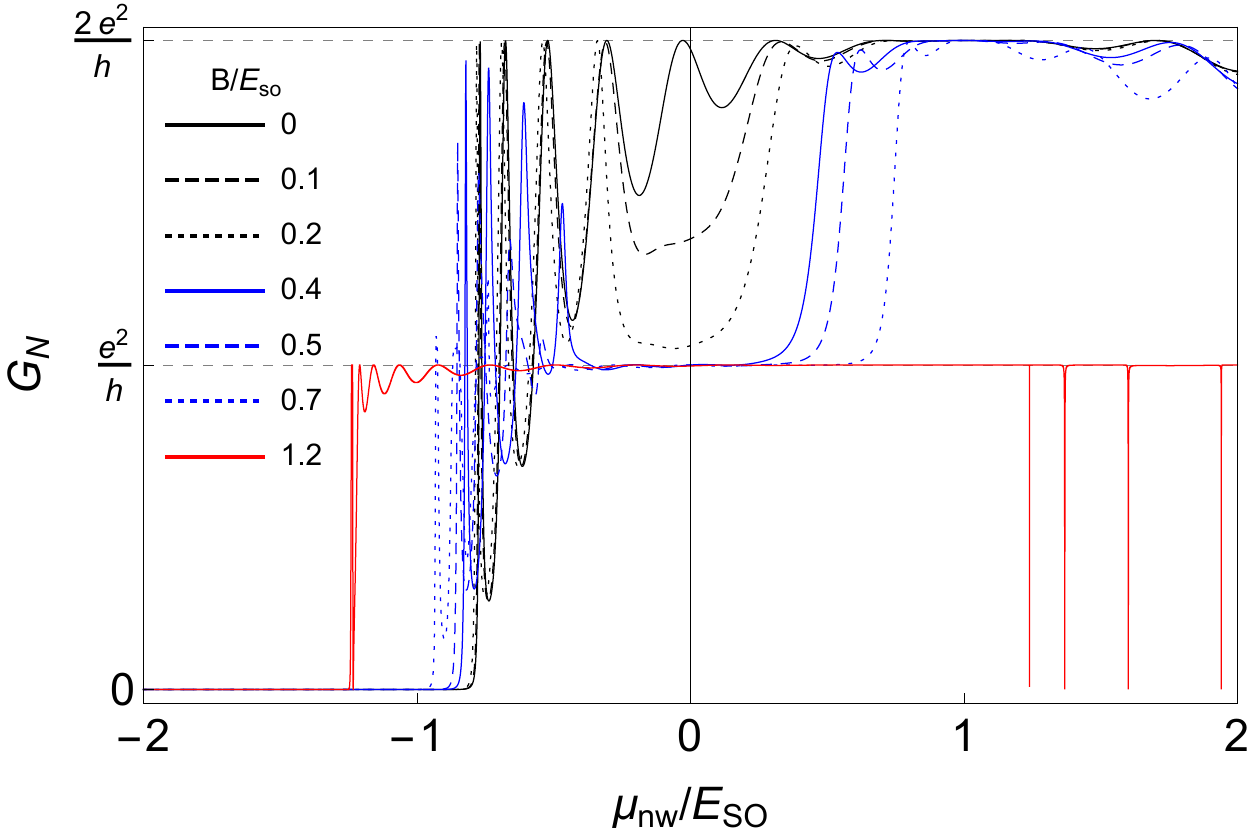} 
\caption{(Color online) Normal conductance $G_{N}$ as a function of the Fermi energy $\mu_{\mathrm{nw}}$ for a long junction with $L_{nw}=4\mu m$,  $E_{SO}=0.05$\,meV and $\mu_{\mathrm{leads}}=E_{SO}$. For intermediate magnetic fields, $B\leq E_{SO}$ the conductance develops a clear helical gap inside the Fabry-Perot resonant structure. This gap signals the region where the middle section of the NW becomes helical. When $B\geq \mu_{\mathrm{leads}}$, the contacts become helical too and the conductance shows helical Fano dips \new{(red curve)}.}
\label{fig4}
%\end{minipage}
\end{figure}
 \begin{figure}[!ht]
%\begin{minipage}[t]{\linewidth}
\centering
\includegraphics[width=.5\textwidth]{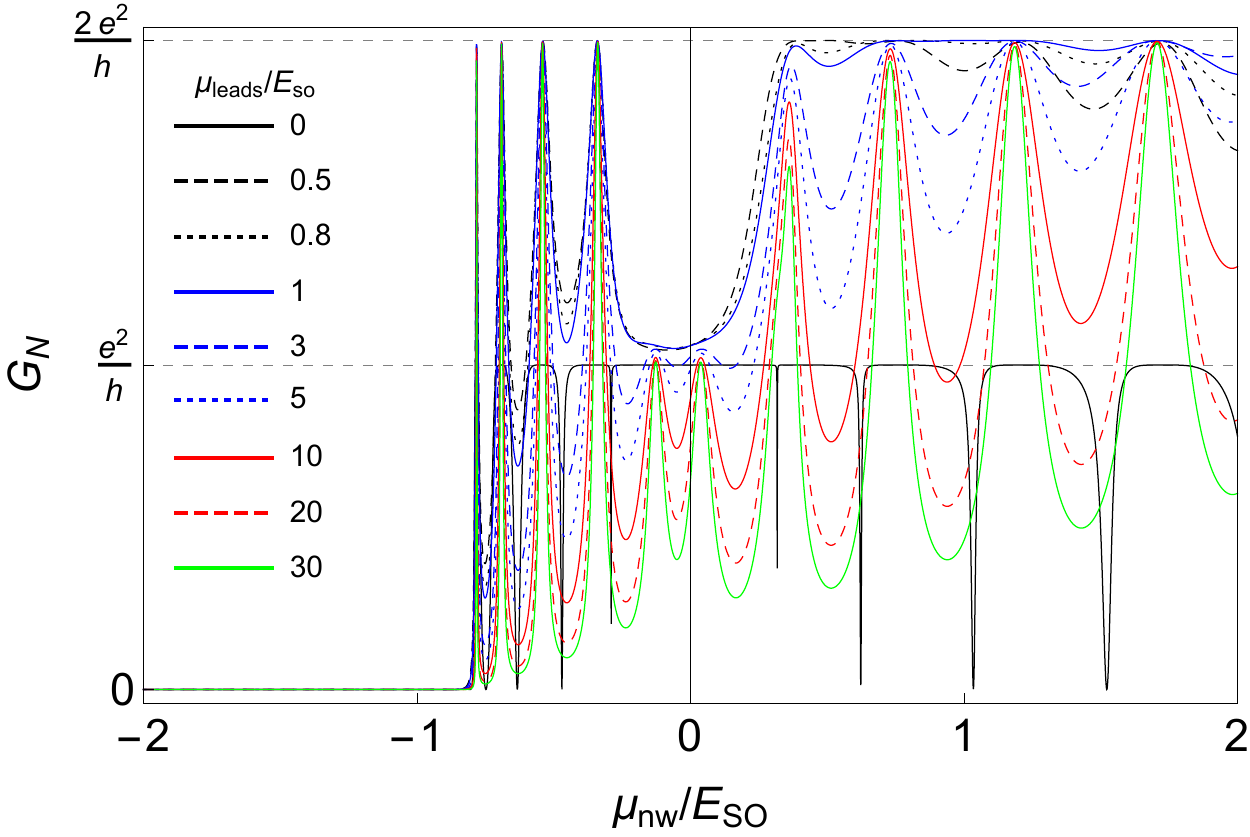} 
\caption{(Color online) Same as Fig.\,\ref{fig4} for fixed magnetic field $B=0.2E_{SO}$ and increasing $\mu_{\mathrm{leads}}$. The helical Fano dips are only seen for $\mu_{\mathrm{leads}}<B$ (solid line).}
\label{fig5}
%\end{minipage}
\end{figure}
 \begin{figure}[!ht]
%\begin{minipage}[t]{\linewidth}
\centering
\includegraphics[width=.5\textwidth]{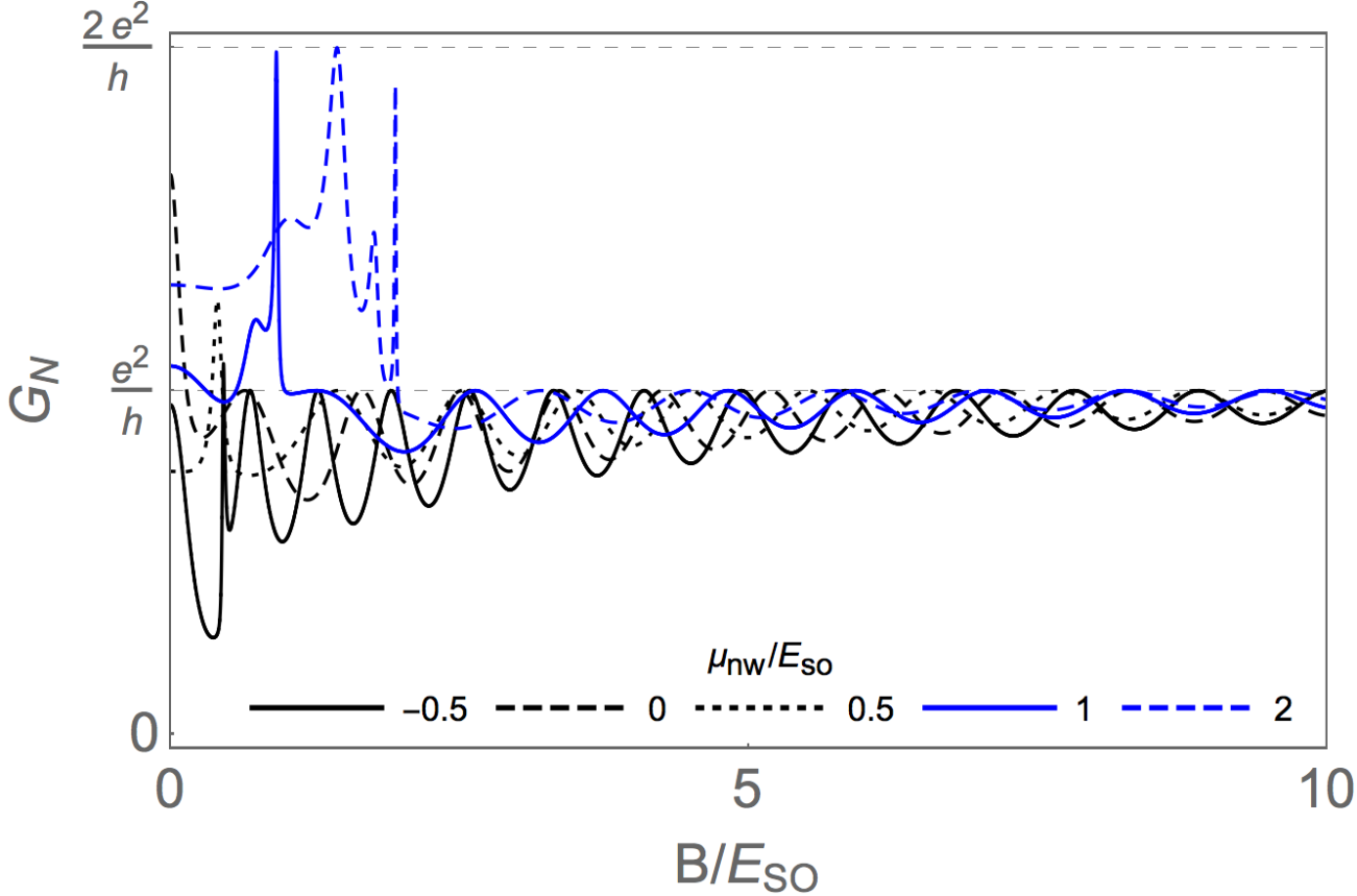} 
\caption{(Color online) Normal conductance $G_{N}$ as a function of magnetic field for different values of the Fermi energy $\mu_{\mathrm{nw}}$ (same parameters as in Fig.\,\ref{fig4}, except $\mu_{\mathrm{leads}}=10E_{SO}$). The oscillatory behavior when $B>\mu_{\mathrm{nw}}$ reflects the transition to the helical regime in the normal side.}
\label{fig6}
%\end{minipage}
\end{figure}

To set the stage, we first \new{consider an NW-N junction between a semi infinite nanowire and a good metal}, which will allow us to discuss deviations when we consider confinement effects. 
Fig.\,\ref{fig1} shows the expected conductance profile as a function of the NW Fermi energy $\mu_{\mathrm{nw}}$, \new{for different values $\mu_{\mathrm{lead}}$ of the Fermi energy in the metal}. At finite magnetic fields, the normal conductance exhibits a gap (with $G_N\approx e^2/h$) of size $\Delta\mu_{\mathrm{nw}}=2B$. As we explained, this gap is a direct consequence of the combined action of Zeeman effect and strong SO coupling and reflects the presence of helical transport, namely spin-polarized counter propagating states \cite{helical}. As discussed in Ref. \onlinecite{PhysRevB.90.235415}, the visibility of this helical gap depends on various factors which, importantly, include the actual value of the SO energy. Indeed, as the ratio $\mu_{\mathrm{lead}}/E_{SO}$ is made larger, the visibility of the gap in $G_N$ is rapidly degraded \new{(see lower curves in Fig.\,\ref{fig1})}.

We now consider the confined N-NW-N junction geometry. \new{Due to the confinement of the central NW section, Fabry-Perot resonances are expected.} Fig.\,\ref{fig2} shows the extreme case of a very short central region with only one resonant \new{quasibound state in the junction}. As expected, the conductance without external Zeeman field \new{(solid curve)} has a Lorentzian shape and reaches its maximum value $G_{N}=2e^2/h$ when $\mu_{\mathrm{nw}}=\mu_{\mathrm{leads}}$ \new{(vertical dashed guideline)}. Similar results are found for small Zeeman fields $B<\mu_{\mathrm{leads}}$ (dashed). When $B=\mu_{\mathrm{leads}}$, \new{however,} the leads becomes spin-polarized \new{(or helical, to be precise)} and hence the maximum conductance is halved to $G_{N}=e^2/h$ (red curve). 

\new{We consider first the situation with $B>\mu_{\mathrm{leads}}$. This regime is} characterised by strong Fano dips that appear when $\mu_{\mathrm{nw}}$ is positive, namely when the junction is gated to create a quantum dot instead of a barrier, see Eq. (\ref{Leq1}). At these Fano dips destructive interference is maximum and $G_{N}=0$. Moreover, the position of these Fano resonances moves to higher $\mu_{\mathrm{nw}}$ as $B$ increases (Fig.\,\ref{fig2}, insets).
The Fano dips can be understood by noticing that the system develops a \new{truly bound state at an energy below $\mu_{\mathrm{leads}}$ as $\mu_{\mathrm{nw}}$ increases (Fig.\,\ref{fig3}a). While for $B\ll\mu_{\mathrm{leads}}$ this level lies far below the chemical potential of the leads and cannot significantly affect $G_N$, in the case $B>\mu_{\mathrm{leads}}$ at hand, the situation is markedly different. At such high fields, one spin sector in the leads is removed away from the chemical potential, and the leads become helical. Similarly, the bound state below $\mu_{\mathrm{leads}}$ is Zeeman-split, such that the component corresponding to the removed spin sector may then cross the Fermi level at a given $\mu_{\mathrm{nw}}$ (Fig.\,\ref{fig3}b-d).  
%two different physical situations may happen: 1) When $B\ll\mu_{\mathrm{leads}}$, both spin projections of the Zeeman-split bound state are well below \new{the chemical potential of the lead} for all $\mu_{\mathrm{nw}}$, so they cannot affect $G_N$ significantly. 2)  \new{On the contrary, when $B\geq\mu_{\mathrm{leads}}$ one spin sector in the leads is removed away from the chemical potential, and the leads become helical. The bound state from the removed sector may then cross the Fermi level at a given $\mu_{\mathrm{nw}}$ (Fig.\,\ref{fig3}b-d). 
%Above this $B$, such situation is always possible (Figs.\,\ref{fig3}c, \ref{fig3}d) but, importantly, the continuum at the contact is now helical. 
This results in one spin projection strongly coupled to the continuum (the sector that is not removed), while the other spin projection remains weakly coupled to this helical continuum through the split bound state (dashed circles), owing to the small spin canting induced by SO}. This configuration mimics the physics of a Fano resonance, as we explicitly demonstrate in Appendix  \ref{tightc} with an effective model.  Note that SO is essential to mimic the physics of the Fano effect (two channels with very different coupling to the continuum). Indeed, we have checked that for $\alpha_R=0$ (namely a fully spin-polarized system without canting) the effect disappears (not shown). The general behaviour is related to the so-called Fano-Rashba effect in systems with inhomogeneous Rashba couplings \cite{PhysRevB.74.153313,Fano-Rashba-B} although in our case the bound states originate from the Fermi energy inhomogeneity, which is probably more realistic for NWs with gates. For intermediate lengths, the system can accommodate many of the above resonances but the helical gap is not visible (not shown). %In the region $\mu_{\mathrm{nw}}<\mu_{\mathrm{leads}}$  (namely, when the normal region is effectively a barrier) these Fabry-Perot resonances are spin-polarized (as it is clear from the conductance value $e^2/h$, arrows in Fig.\,\ref{fig4}) and reflect transport across the helical normal region. In the region $\mu_{\mathrm{nw}}>\mu_{\mathrm{leads}}$, the conductance also develops helical Fano dips. 

\new{Consider now the $B<\mu_{\mathrm{leads}}$ regime complementary to the preceding discussion. In this situation, there exist two propagating channels in the leads, and conductance may reach $2e^2/h$ at Fabry-Perot maxima, as long as the central NW is likewise non-helical ($B>|\mu_{\mathrm{nw}}|$). Otherwise, for long enough junctions ($L_{\mathrm{nw}}\geq 4\mathrm{\mu m}$ for the realistic NW parameters in our simulation) a helical gap develops in conductance, such that $G_N\lesssim e^2/h$. As central $\mu_{\mathrm{nw}}$ is tuned into and out of the helical regime, conductance exhibits a reentrant behavior, switching from $\sim 2e^2/h$ to $e^2/h$ and back to $2e^2/h$. This reentrance can be resolved across multiple resonant helical Fabry-Perot oscillations.
This is illustrated in Fig.\,\ref{fig4} where we plot the conductance for a 4$\mathrm{\mu m}$-long nanowire as a function of the central Fermi energy $\mu_{\mathrm{nw}}$. Note the reentrant conductance, and the helical Fabry-Perot resonances with an $e^2/h$ ceiling, signalling helical transport in the junction.}
%As expected, the conductance gap signalling helical behavior in the middle part of the NW (with a reentrant behavior of conductance from $\sim 2e^2/h$ to $e^2/h$ and back to $2e^2/h$) 
\new{The visibility of the conductance reentrance and the helical gap is lost for fields $B>E_{SO}$, see bandstructure inset in Fig. \,\ref{fig1}. At such fields, the helical gap becomes an extended $G_N\sim e^2/h$ half-plateau (potentially with superimposed Fano resonances if $B$ also exceeds $\mu_{\mathrm{leads}}$) that emerges directly from pinch-off $G_N=0$.} 
%When the NW sections below the contacts become helical for $B\geq\mu_{\mathrm{leads}}$, the conductance also shows helical Fano dips superimposed to the half-plateau. 
Note that the regime with helical Fano dips in the normal conductance is quite relevant towards reaching topological superconductivity: the NW under the contacts can become a non-trivial topological superconductor in the presence of pairing as long as it can be depleted and made helical in the normal phase. Hence our prediction of helical Fano dips superimposed on a half-plateau of $G_N\sim e^2/h$ constitutes a strong signature of helical behaviour as precursor of non-trivial superconductivity. 

Similar phenomenology is obtained for conductance at fixed magnetic fields and increasing $\mu_{\mathrm{leads}}$ (Fig.\,\ref{fig5}). As expected, the Fano dips disappear as soon as $\mu_{\mathrm{leads}}>B$ while the gap coming from helicity in the central section in the NW is much more robust. Increasing $\mu_{\mathrm{leads}}$ results in well defined Fabry-Perot resonances in the helical gap region. The normal conductance as a function of magnetic field is shown in Fig.\,\ref{fig6}. Here, a change from irregular behavior to regular $e^2/h$ oscillations as a function of magnetic field signals the helical regime when $B\geq\mu_{\mathrm{nw}}$ \cite{PhysRevB.90.235415}. 

%We emphasize that the resonant effects discussed here, while originating from helicity, are different from the gaps in the conductance of a semi-infinite N-NW junction that we discussed at the beginning of the section and in Fig.\,\ref{fig1}. 
Having in mind that there exists no conclusive experimental evidence of the helical regime in nanowires in the literature\cite{nphys610,VanWeperen}, the nontrivial resonant effects in finite-length junctions that we have described, both helical Fabry-Perot resonances and helical Fano dips, could be used as an interesting option for \new{detecting} such helical transport regime in long junctions. \new{Even more significant}, these helical resonant features give rise to a non-trivial subgap spectrum when the leads become superconducting, as we discuss in what follows.
\begin{figure}[!ht]
\centering
\includegraphics[width=.45\textwidth,height=.25\textwidth]{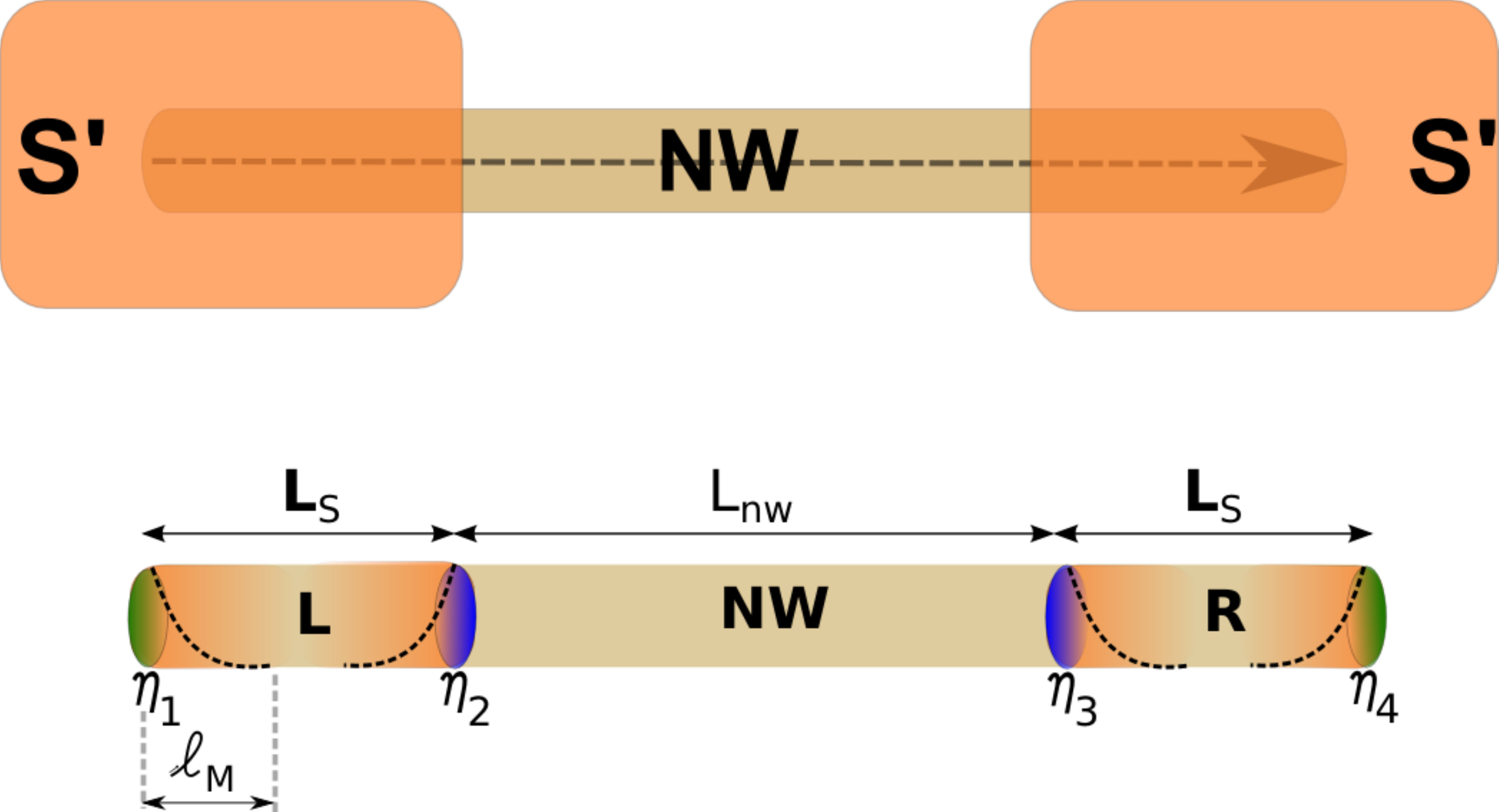} 
\caption{(Color online) Two s-wave superconducting contacts (S', with gaps $\Delta_{S'}$) deposited on top of a Rashba nanowire (NW) of length $L=L_{S}+L_{\mathrm{nw}}+L_{S}$. The superconductors induce superconducting correlations into some regions of the nanowire via proximity effect, giving rise to regions which we refer to as superconducting leads (left L and right R) with gaps $\Delta<\Delta_{S'}$ and Fermi energies $\mu_{\mathrm{leads}}$, and a central region in the normal state with $\mu_{\mathrm{nw}}$. The dashed arrow in the first figure denotes the applied Zeeman field along the NW. Due to the finite length $L_S$, the junction in the topological phase hosts four Majorana bound states, $\eta_{1}, \eta_{2}, \eta_{3}, \eta_{4}$, for a phase difference of $\pi$ between the superconductors, with localisation length  $\ell_{M}$.}
\label{fig7}
\end{figure}

\begin{figure*}[!ht]
\centering
\includegraphics[width=0.7\textwidth]{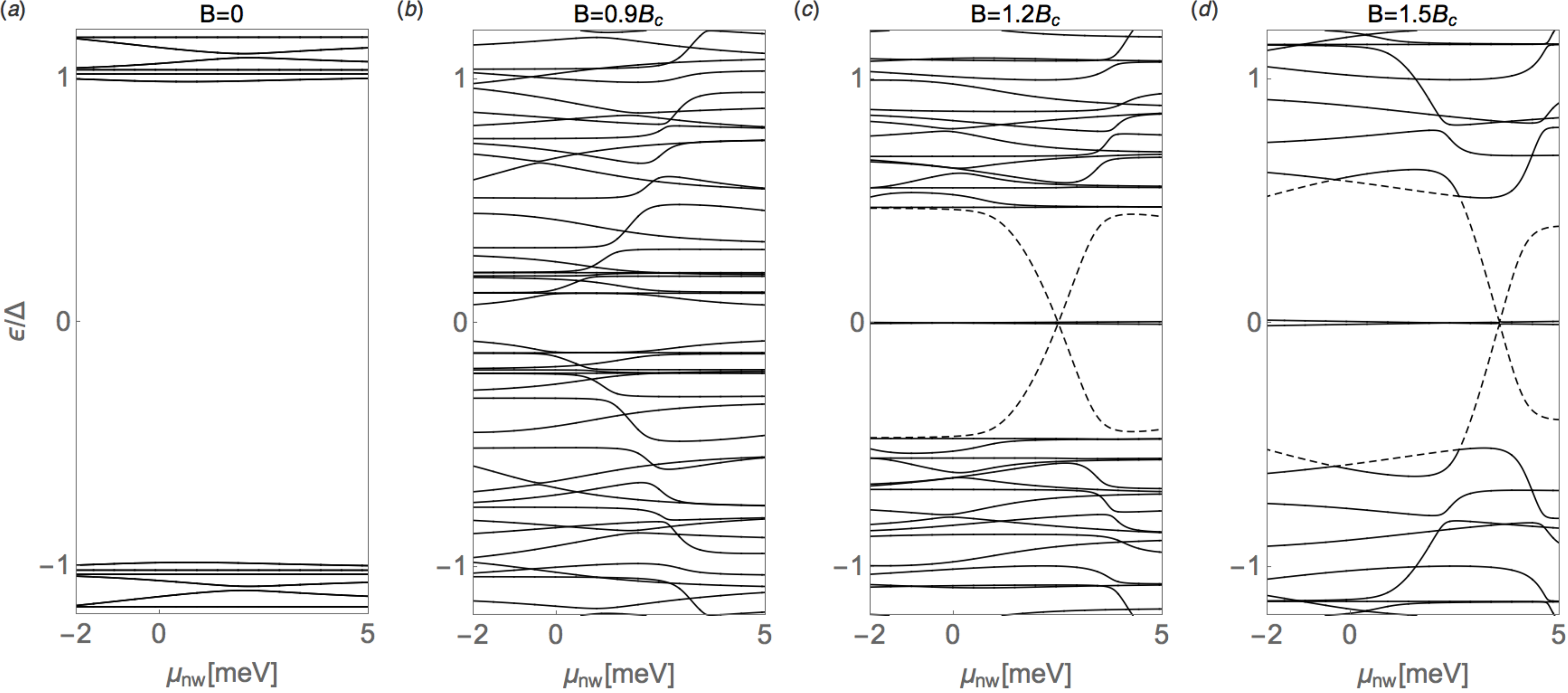} 
\caption{Andreev levels  at $\varphi=0$ of a short junction, $L_{\mathrm{nw}}=20$\,nm as a function of  $\mu_{\mathrm{nw}}$. Different panels show the evolution of the spectrum for increasing magnetic fields. Parameters: $E_{SO}=0.05$\,meV, $\mu_{\mathrm{leads}}=10E_{SO}$, $L_{S}=2\mu$m, $\Delta=0.25$\,meV.}
\label{fig8}
\end{figure*}

\section{Subgap levels in SNS junctions}
\label{ElevelsSNS}
\subsection{SNS junction model and relevant length scales}
To model a SNS junction we assume that the outer parts of the wire are coupled to an s-wave superconductor (with bulk values $\mu_{S'}$ and pairing $\Delta_{S'}$), while the central is not (see Fig.\,\ref{fig7}).
%We consider two s-wave superconductors (S') with chemical potential $\mu_{S'}$ and pairing $\Delta_{S'}$, in thermal equilibrium, linked by a nanowire with Rashba spin-orbit coupling and Zeeman effect along it, Fig.\,\ref{fig7}. 
Superconducting correlations are induced by proximity effect into the nanowire. For good enough contact between the NW and the superconductor, the Cooper pair amplitude is finite inside the NW regions below the superconductor. In most papers in the literature, this situation is modeled by including by hand a pairing term, $\Delta<\Delta_{S'}$, in the hamiltonian of such NW regions.  While, rigorously speaking, this is incorrect (the superconducting coupling constant is zero inside the NW), it is well known that it provides a good description of the proximity effect for large enough gaps (in such cases, the parameter $\Delta$ is essentially the low frequency limit of a tunneling self-energy and is given by the tunnel coupling between the normal and superconducting parts, see e.g. \onlinecite{Bena2}). Therefore, we adopt this approximation here for simplicity (we have checked that all our conclusions remain unaltered irrespective of whether we use this simplified model or a full NW + SC coupling model, see appendix A.3). In cases where the interface transparencies are small, extra Fabry-Perot resonances coming from insulating layers could complicate our analysis, see Ref. \onlinecite{Galaktionov-Zaikin}. 

In particular, we model the regions of the nanowire below the superconducting contacts as regions with Fermi energy $\mu_{leads}$ and \new{pairing potential on the left (L) and right (R) contact given by} $\Delta_{L}=\Delta\,{\rm e}^{- i \varphi/2}$ and $\Delta_{R}=\Delta\,{\rm e}^{ i \varphi/2}$, with $\Delta<\Delta_{S'}$.
The region in the middle of the nanowire without superconducting correlations is the normal region (N) with Fermi energy \new{denoted by $\mu_{\mathrm{nw}}$ as before}. At high enough magnetic fields, the regions of the NW below the superconductors (S regions of the junction) can be driven into a topological superconducting phase when \new{$B>B_{c}\equiv\sqrt{\mu_S^2+\Delta^2}$}. Owing to the finite length $L_S$, this results in a SNS junction with four Majorana bound states for a phase difference of $\pi$ between the
superconductors: two inner Majorana bound states, labeled $\eta_{2,3}$, that form inside the junction, and two outer Majorana bound states, $\eta_{1,4}$, see Fig.\,\ref{fig7}. On the other hand, for a zero phase difference, only the outer MBSs are present.

SNS Josephson junctions are classified in two types, depending on the relationship between the length of the normal region $L_{\mathrm{nw}}$ (i.e. distance between the superconducting \new{contacts}) and the coherence length $\xi=2\hbar v_{F}/\pi\Delta$, where $v_{F}$ is the Fermi velocity. Short junctions are characterized by $L_{\mathrm{nw}}\ll\xi$, whereas $L_{\mathrm{nw}}\gg\xi$ in long junctions. Such classification can be also given in terms of natural energy scales of the problem, the Thouless energy, $E_{T}=\hbar v_{F}/L_{\mathrm{nw}}$, and the induced superconducting pair potential $\Delta$, being $v_{F}$ the Fermi velocity, and $L_{\mathrm{nw}}$ the length of the normal region. %The Thouless energy is a single-electron quantity that represents the diffusion rate for a single electron across the normal region of length $L_{\mathrm{nw}}$, while the pair potential $\Delta$ is set by the interactions in the superconducting electrodes. 
The above conditions, in terms of these energy scales, are $\Delta\ll E_{T}$ for short junctions and $\Delta\gg E_{T}$ for long ones. \new{The significance of this classification is related to the typical number $\sim \Delta/E_{T}$ of Andreev subgap states of the junction, in addition to the MBSs at zero energy.}

The MBSs wave functions decay from both ends of the topological superconducting leads.  The inner and outer MBSs may feel their mutual presence if their wave functions exhibit a non zero spacial overlap. The relevant decay distance characterizing this overlap is the Majorana localization length $\ell_{M}$ (appendix \ref{Majorana-length}). 
For finite $L_{S}<2\ell_{M}$ the overlap between MBSs is significant and therefore they are no longer true zero modes.

In what follows, we discuss the subgap spectrum of short SNS junctions in the topological regime $B>B_c$ as well as the subgap spectrum of long SNS junctions as one goes from the \emph{helical} junction regime to the topological one. \new{The helical junction regime is defined by a central region depleted into the helical regime, while the S regions remain non-topological, namely by $\mu_{leads}>\mu_{\mathrm{nw}}$, and $\mu_{\mathrm{nw}}<B<B_c$.}

\subsection{Short junctions}
\label{short}
For very short junctions, the ABS spectrum at $B<B_c$ and $\varphi=0$ does not contain sub-gap states (Figs. \ref{fig8}a and b). This is expected for a short junction with $\xi\gg L_{nw}$. % and helicity $l_{SO}\gg L_N$ (such a short junction does not contain helical Fabry-Perot resonances, see Fig. \ref{fig2}). 
The $B>B_c$ spectrum (Figs. \ref{fig8}c and d), on the other hand, \new{is much more interesting.} It contains the expected subgap state near zero energy for all $\mu_{\mathrm{nw}}$ (coming from the weakly coupled outer Majoranas for $L_S>\ell_{M}$, the inner MBS at $\varphi=0$ are strongly hybridized and form standard ABS at energy $\sim\Delta$). Notably, this MBS coexists  with a bound state that crosses zero energy for a given $\mu_{\mathrm{nw}}>0$ (dashed line). This bound state originates from the single resonance that the junction accommodates for increasing $\mu_{\mathrm{nw}}>0$ (see Fig. \ref{fig3}), \new{which we discussed in connection to Fano resonances}. If we interpret this resonant state as an impurity level, our results for $B<B_c$ are consistent with Anderson's theorem which prevents the existence of bound states inside the gap of an s-wave superconductor for non-magnetic impurities \cite{Anderson-theorem}. \new{The reason is that the zero-enery crossing appears for $B>B_c$, such that the superconductor is effectively p-wave}. Therefore, the emergence of these subgap states crossing zero energy should be understood as a direct consequence of nontrivial topology in the junction \cite{Sau-Demler,Huetal}. The precise condition for the level crossing coincides with the condition for having a Fano dip. As we discussed in section \ref{normCon}, this is the condition in the normal regime for having a single resonant state which interferes destructively with a helical contact; the latter condition is here fulfilled because $\mu_{\mathrm{leads}}<B_c<B$. These subgap states and zero-energy crossings should be understood as the p-wave counterparts of so-called Yu-Shiba-Rusinov sub-gap states \cite{subgap1,subgap2,subgap3,subgap4} and their corresponding parity crossings  \cite{subgap5} in s-wave superconductors with magnetic impurities. %Parity crossings of nonmagnetic origin appear here owing to the effective p-wave pairing in the topological phase. %Therefore, the emergence of these subgap states crossing zero energy should be understood as a direct consequence of nontrivial topology in the junction \cite{Sau-Demler,Huetal}. %This is demonstrated in the central panel where the plot the spectrum at $B=0.9B_c$: although the system in the normal regime already contains helical resonances for $\mu_{\mathrm{nw}}<0$, the corresponding superconducting junction does not support parity crossings for $B<B_c$.

%Note that the dependence with $\mu_{\mathrm{nw}}$ of the bound states in the topological phase is quite different from the one expected from magnetic Yu-Shiba-Rusinov subgap states, like the ones appearing in quantum dots in the Coulomb Blockade regime. This difference would allow in principle to rule out zero-bias anomalies coming from interaction effects  \cite{Lee:13,Zitkoetal} in accidentally-formed quantum dots inside the NW.
\begin{figure}[!ht]
\centering
\includegraphics[width=0.4\textwidth,height=.25\textwidth]{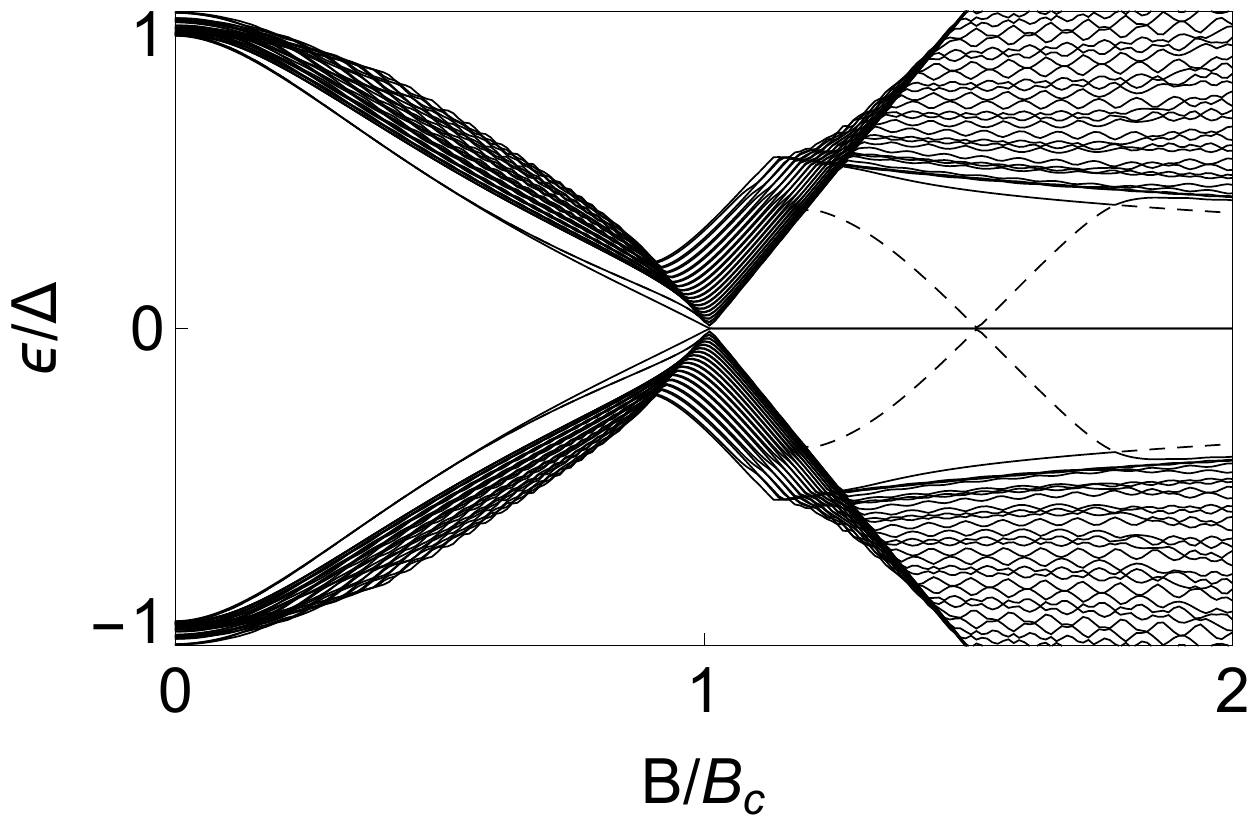} 
\caption{Andreev levels at $\varphi=0$ as function of the Zeeman field for $\mu_{\mathrm{nw}}=3.57$\,meV.  The rest of parameters are the same as in Fig. \ref{fig8} except $L_S=10\mu m$.}
\label{fig9}
\end{figure}

 \begin{figure*}[!ht]
\centering
\includegraphics[width=.7\textwidth]{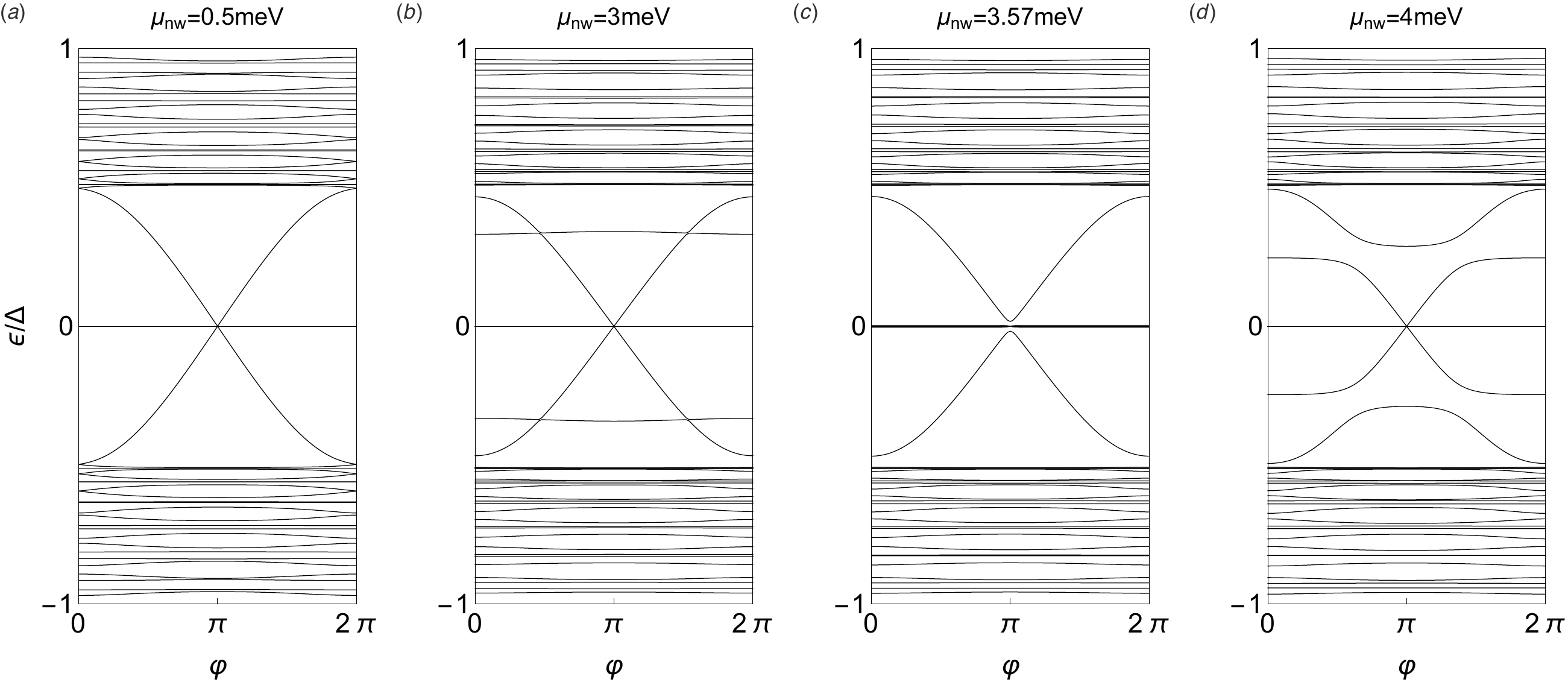} 
\caption{Andreev levels at the junction $\epsilon(\varphi)$ in the short-junction regime, $L_{\mathrm{nw}}=20$\,nm in at $B=1.5B_{c}$. Parameters: $\alpha_{R}=20$\,meV\,nm for InSb nano wires, $\mu_{leads}=0.5$\,meV, $L_{S}=10\mu m$, and $\Delta=0.25$\,meV. Different panels show the Andreev levels around $\mu_{\mathrm{nw}}=3.57$\,meV near the zero-energy crossing in Fig.\,\ref{fig8}d.}
\label{fig10}
\end{figure*}

\begin{figure*}[!ht]
\centering
\includegraphics[width=\textwidth]{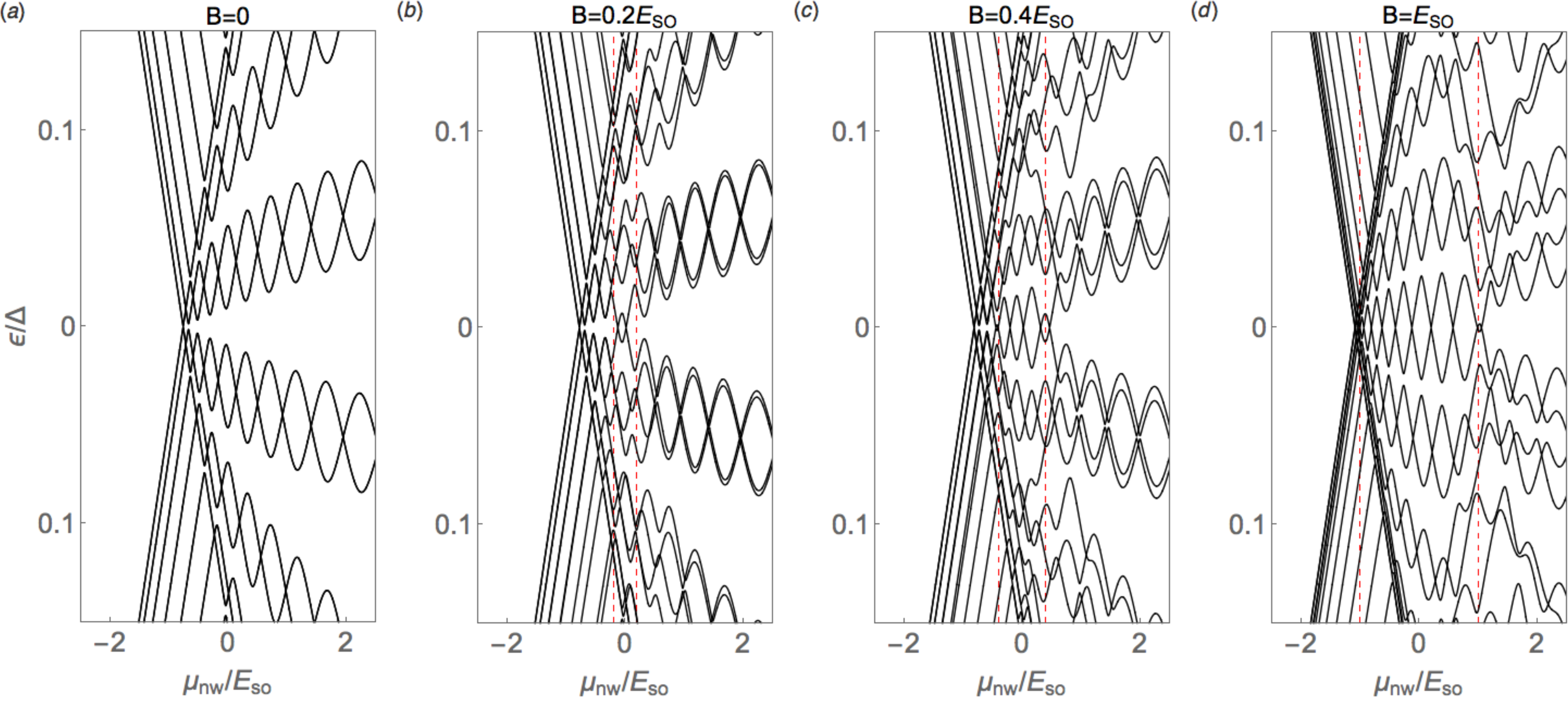} 
\caption{Andreev levels at $\varphi=0$ as function of $\mu_{\mathrm{nw}}$ for a long junction, $L_{\mathrm{nw}}=4\mu$m and various magnetic fields. Parameters: $E_{SO}=0.05$\,meV, $\mu_{leads}=10E_{SO}$, $L_{S}=2\mu$m, $\Delta=0.25$\,meV. At finite B, the ABS spectrum shows a loop structure around zero energy in the region where the normal side becomes helical (marked by dashed lines). Note that the junction is very far from becoming topological ($B_c\approx 11.2 E_{SO}$).}
\label{fig11}
\end{figure*}  
Further insight comes from the magnetic field dependence at fixed $\mu_{\mathrm{nw}}$ (Fig. \ref{fig9}). After the closing of the gap across the topological phase transition at $B=B_c$, the spectrum of the junction exhibits  a perfect zero-energy state accompanied by a zero-energy crossing  (dashed line) similar to the one discussed in Fig. \ref{fig8}. %We emphasize that this Shiba state is absent for $\mu_{\mathrm{nw}}=\mu_S$ (not shown). 
Note here that, despite the finite length of the central NW, the zero energy state for $B>B_c$ does not oscillate as a function of Zeeman field, \new{unlike what is typical of overlapping MBSs} \cite{Lim:PRB12,Prada:PRB12,Rainis:PRB13,DasSarma:PRB12}. This can be easily understood as this state comes from the \emph{outer} MBSs which at $\varphi=0$ are effectively decoupled across the junction, \new{since we assume $L_S\gg \ell_M$. }

We now analyse in more detail the full phase dependence in the topological phase for different values of $\mu_{\mathrm{nw}}$. The low-energy sector is characteristic of a short junction: two almost $\varphi$-independent levels near zero energy coming from outer MBSs and two dispersive levels coming from hybridization of inner MBSs across the junction. The anti crossings near $\varphi=\pi$ are only visible for finite $L_S/\ell_{M}$. For $L_S\gg \ell_{M}$ (Fig.  \ref{fig10}a), the zero-energy levels are flat and the anti crossing at $\varphi=\pi$ becomes negligible \footnote{In $L_S\rightarrow\infty$ limit, the outer Majoranas \new{are no longer involved in transport} while the levels at $\varphi=\pi$ exactly cross (not shown) giving rise to anomalous $4\pi$-periodic spectrum and Josephson currents if fermionic parity is conserved}. In the following, we refer to the dispersive ABS with almost perfect crossings at $\varphi=\pi$ as Majorana ABSs. As $\mu_{\mathrm{nw}}$ increases, an extra bound state emerges from the continuum as an almost dispersionless subgap state and interacts very weakly with the Majorana ABSs (Fig.  \ref{fig10}b). Importantly, after crossing zero energy (Fig.  \ref{fig10}c) and reemerging at finite energy (Fig.  \ref{fig10}d), the anti crossing with the Majorana ABS is considerably larger, \new{indicating} that the bound state has changed its parity character.

\subsection{Long junctions}
\label{long}
The ABS spectrum of long junctions at small magnetic fields $B<B_c$ differs considerably from the one of short junctions. Even for $B=0$ (Fig.  \ref{fig11}a), the spectrum is very sensitive to the sharp increase of conductance at small negative $\mu_{\mathrm{nw}}$, when the junction goes rapidly from pinch-off to fully transmitting (solid \new{black line} in Fig.\,\ref{fig4}). This is reflected in a feature that resembles the closing and reopening of a gap (but, of course, \new{is  related to the central region becoming metallic, rather than with} a gap closing). The emergence of Fabry-Perot resonances in the normal phase is translated into the appearance of level pairs at finite energies, or loops, that oscillate with system parameters in the superconducting phase. \new{A distinct change in the loop structure takes place as $B$ is increased within a window $|\mu_{\mathrm{nw}}|<B$. This, recall, corresponds to the helical regime of the normal region, characterised in normal transport by a helical gap and helical Fabry-Perot oscillations. The loops inside said window reconnect, and give rise to new loops around zero-energy, separated by parity crossings (Fig.  \ref{fig11}b)}. Each of these crossings corresponds to a helical Fabry-Perot resonance in the normal regime.
% Importantly, the loops around zero energy appear in regions of $\mu_{\mathrm{nw}}$ where the \emph{normal side} of the junction becomes helical (dashed lines). 
%Therefore, an increase in the Zeeman energy, should result in a sizeable range of $\mu_{\mathrm{nw}}$ 
\new{For larger Zeeman energies, supporting many helical Fabry-Perot resonances within the helical gap, correspondingly many consecutive zero-energy loops become visible in the superconducting regime}. As soon as the normal side ceases to be helical ($|\mu_{\mathrm{nw}}|>B$), the spectrum does no longer show loops around zero energy. Since depleting the normal section of the NW junction should be much easier than gating the proximized region, we expect that said near-zero loops and parity crossings should be ubiquitous for finite size junctions near depletion \footnote{intermediate $L_{\mathrm{nw}}$ junctions also show the same behaviour, not shown} and \new{constitute yet another alternative scheme to detect} the helical regime. 

\begin{figure}[!ht]
\centering
\includegraphics[width=.5\textwidth]{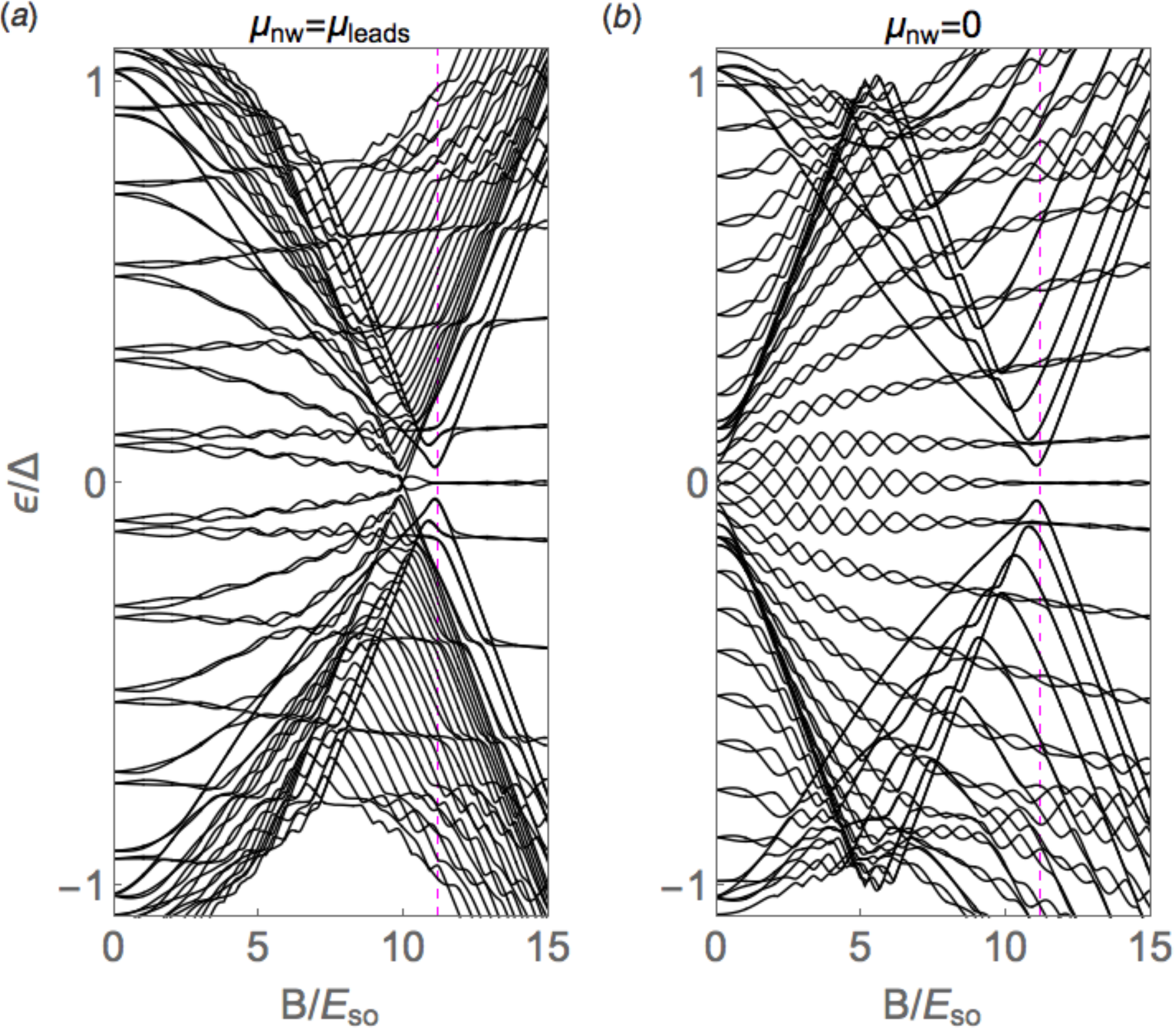} 
\caption{Andreev levels at $\varphi=0$ as function of $B$. Same parameters as in Fig.\,\ref{fig11}. \new{The critical field $B_c$ is marked by vertical dashed line.}}
\label{fig12}
\end{figure}
\begin{figure}[!ht]
\centering
\includegraphics[width=.5\textwidth]{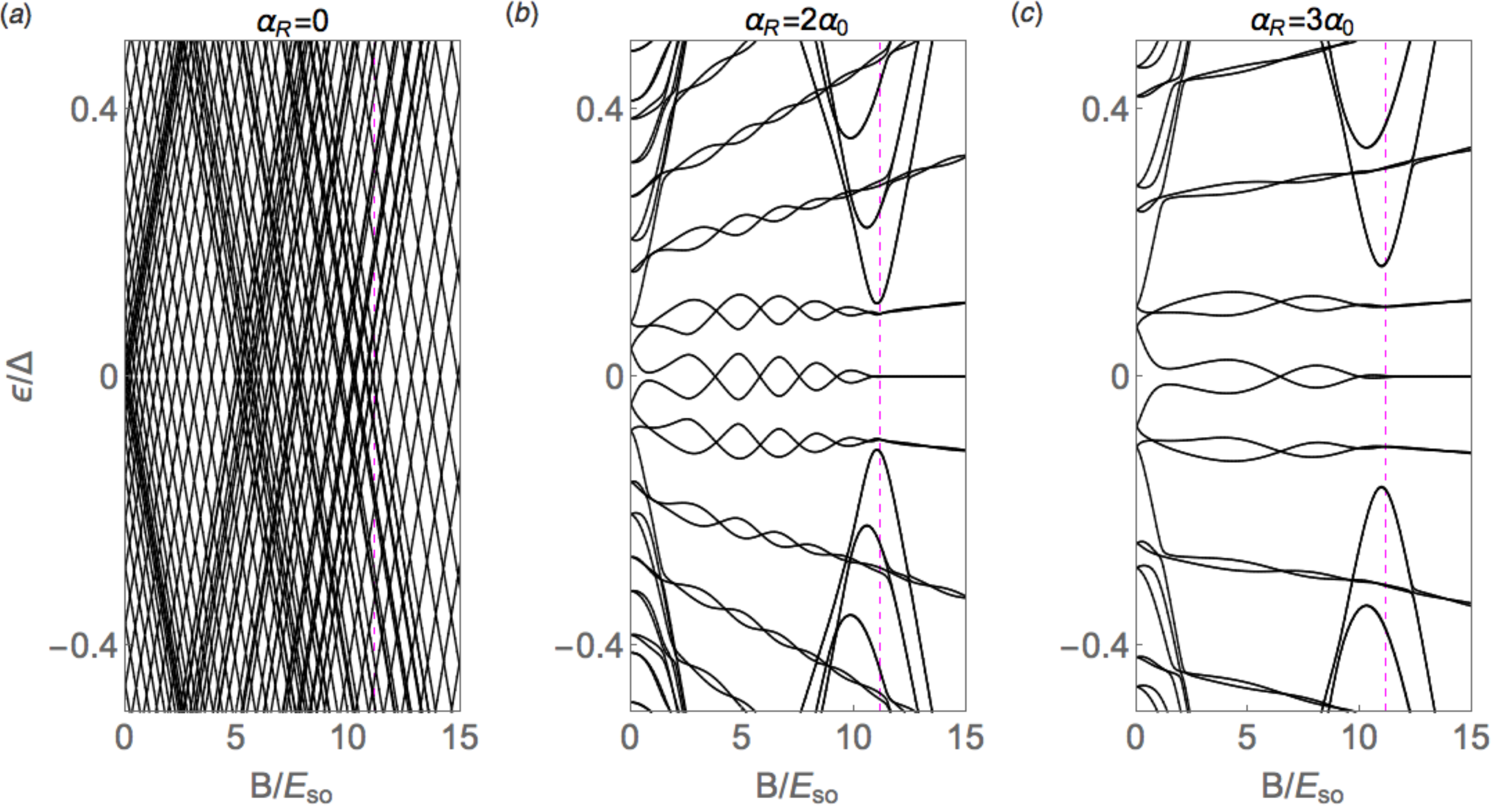} 
\caption{Same as Fig.\,\ref{fig12} for $\mu_\mathrm{nw}=0$ and increasing values of the SO coupling $\alpha_R$. \new{The critical field $B_c$ is marked by vertical dashed line.}}
\label{fig13}
\end{figure}
Each loop in the helical regime (see e.g. Fig.  \ref{fig11}b) is similar to the ones expected for magnetic impurities \cite{subgap1,subgap2,subgap3,subgap4}, or quantum dots in the Coulomb blockade regime \cite{Lee:13,PhysRevLett.110.217005} coupled to superconductors (we emphasize here that our junction is noninteracting). This result again suggests an interesting analogy with the physics of Yu-Shiba-Rusinov states in superconductors with magnetic impurities. Here, the combined action of Zeeman-induced spin-polarization \emph{and depletion} is crucial. 

\new{Consecutive loops around zero energy, resemble the oscillatory behavior expected from overlapping MBSs} in finite length NWs. \new{However, since the helical gap condition $|\mu_{\mathrm{nw}}|<B$ does not involve $\mu_S$, which may be large, the zero-energy loops may exist while the proximized S regions are still in the topologically trivial regime $B<B_c$ (Fig. \ref{fig11} c and d). Remarkably, there exists a profound connection between zero-loops and MBSs. We find that the former actually evolve continuously into outer MBSs as $B$ is increased beyond $B_c$}.  To illustrate this key idea, we compare in  Fig.  \ref{fig12}, a situation without near-zero energy loops at low B fields ($\mu_{\mathrm{nw}}=\mu_{leads}$, panel a) with another with loops at very low B coming from a helical normal region ($\mu_{\mathrm{nw}}=0$,  panel b). While the MBSs in the first configuration emerge from a situation without zero energy states/crossings at low fields, the ones corresponding to the second configuration are clearly evolving from the low B-field loops around zero energy. We emphasize here that both configurations correspond to the same physical nanowire junction with the sole difference of a depletion in the \emph{normal part} of the junction in the second case. Fig.  \ref{fig12} nicely illustrates two of our main results: 1) long loops with parity crossings in the ABS spectrum can be used to identify the helical regime in a Rashba NW and 2) such helical loops, coming from depletion in the \emph{normal} side of the junction, continuously evolve into MBS for large enough magnetic fields.
\begin{figure}[!ht]
\centering
\includegraphics[width=.5\textwidth]{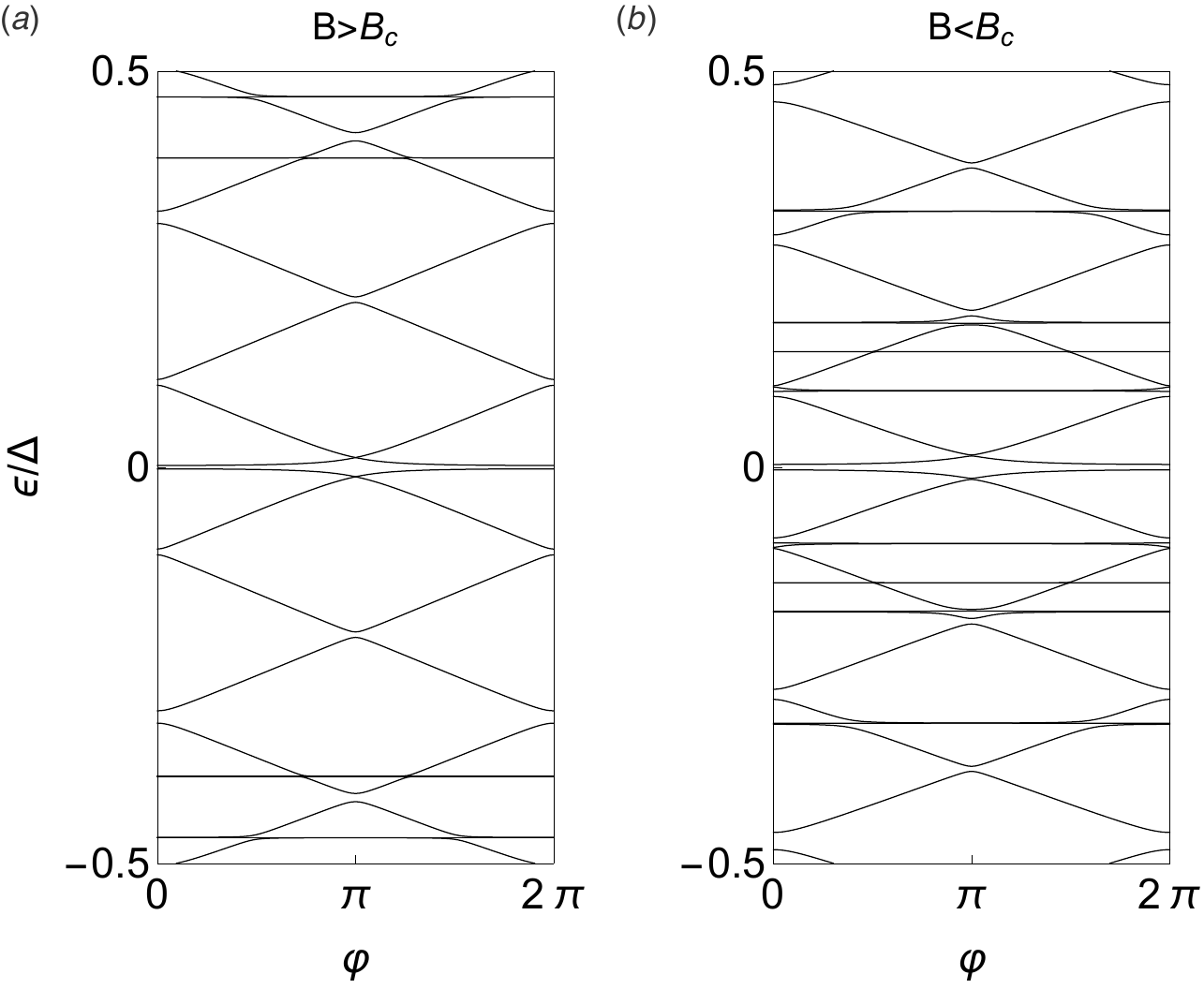} 
\caption{Andreev levels as function of phase $\varphi$ for two values of the Zeeman field a) $B=10E_{SO}$ and b) $B=13E_{SO}$. Rest of parameters same as in Fig. \ref{fig12}b.} 
\label{fig14}
\end{figure}

To obtain more precise information about the nature of this interesting connection between $B<B_c$ near-zero loops and MBS states, we study their evolution for increasing SO coupling (Fig.  \ref{fig13}). For $\alpha_R=0$ (Fig.  \ref{fig13}a), Zeeman-induced depairing closes the superconducting gap and the spectrum becomes a dense quasi-continuum (the full junction is in the normal regime), as expected. Any $\alpha_R\neq 0$ removes all finite energy crossings while preserving the parity-protected crossings at zero energy. As a result, the spectrum is still gapped after the first parity crossing (the Zeeman field is no longer fully depairing) and \emph{many parity crossings} are possible. This important observation is illustrated in Fig.  \ref{fig13}(b,c)  (see also Fig.  \ref{fig12}b). \new{For finite $\alpha_R$}, the low-energy spectrum remains gapped after the first crossing and also after subsequent crossings. Another interesting conclusion that we can draw from our results is that a clear distinction between the near-zero states in the $B<B_c$ and $B>B_c$ regions can no longer be made. The only difference \new{is quantitative, in that} the amplitude of MBS oscillations in the topological regime become smaller for increasing $\alpha_R$, unlike for $B<B_c$. (The SO length becomes much shorter and, hence $L_S\gg\ell_{M}$). \new{However, other spectral properties, such as the mini gap separating the near-zero modes from the first excited states, is roughly the same in both the trivial $B<B_c$ and non-trivial $B>B_c$ phases.}

%We emphasize \new{once more that} the near zero oscillating state can be obtained at very low fields in the topologically trivial region, arbitrarily far from $B_c$. Also worth mentioning is that the mini gap separating the near-zero modes from the first excited states is roughly the same in both $B<B_c$ and $B>B_c$ regions. 

To finish, we consider the phase dependence of the subgap spectrum. While topological SNS junctions with $L_S\rightarrow\infty$ are $4\pi$-periodic as a function of phase difference $\phi$ \new{due to the} characteristic parity-protected crossing at $\varphi=\pi$ (see e.g. Fig.\,\ref{fig10}a), in finite $L_S$ junctions  (Fig. \,\ref{fig14}a), said crossing is avoided, and splits by a small energy due to the hybridization of MBSs at the junction \new{(inner)} and MBSs at the far ends of each S region \new{(outer)}, which leads to a more conventional $2\pi$-periodicity  \cite{San-Jose:PRL12a}. Interestingly, the subgap spectrum at $B<B_c$ (Fig. \,\ref{fig14}b) shows essentially the same phase-dependence which further confirms the deep connection between the $B<B_c$ and $B>B_c$ parity crossings. Note that the resulting Josephson current \cite{Cheng:PRB12}, which only depends on the Andreev spectrum, would be effectively the same (not shown).

\section{Conclusions}
\label{concl}
We have studied the \new{normal transport and the} sub-gap spectrum of SNS junctions based on semiconducting nanowires with strong Rashba spin-orbit coupling. In particular, we have focused on the role of confinement effects in ballistic finite-length junctions and analyzed the distinct properties of the ABS for short and long junctions as different sections of the underlying NW (N or S or both) \new{become} helical. For $B>B_c$, confined levels in the normal section give rise to bound \new{subgap} states, as expected from the effective p-wave nature of the topological superconductor. \new{In normal transport, such bound states give rise to helical Fano dips}. Perhaps more strikingly, we have found that a long junction with a helical normal section, but still in the topologically trivial regime with $\mu_{\mathrm{nw}}<B<B_c$, supports a low-energy subgap spectrum consisting of multiple-loop structures and parity crossings. \new{Such states are derived from helical Fabry-Perot resonances in the normal regime}. We have argued that such multiple loop structure in the ABS spectrum could be used to unambiguously identify the helical regime in NWs. Interestingly, these multiple loops smoothly evolve towards Majorana bound states as the Zeeman field exceeds the critical value. This suggests an interesting connection between subgap parity crossings in helical junctions with $B<B_c$ and Majorana bound states in topological ones with $B>B_c$. A recent study of fully open \new{helical-N/trivial-S} contacts\cite{San-Jose:14b} further confirms the profound connection between subgap states in the helical regime and Majorana physics.

%\begin{figure}[!ht]
%\centering
%\includegraphics[width=.48\textwidth,height=.3\textwidth]{figures/New/long/EmuN_realparam_longjunction_phipi3.pdf} 
%\caption{(Color online) Energy levels as function of the chemical potential in the normal region in the long-junction regime, $L_{N}=1000$\,nm, at $\varphi=\pi$. Left, middle and right panels shows calculations for zero Zeeman, in the helical phase $B=\mu_{leads}$ and the topological phase, respectively.
%Parameters: $\alpha_{R}=20$\,meV\,nm, $L_{S}=2000$\,nm, $\Delta=0.25$\,meV and $\mu_{leads}=0.5$\,meV.}
%\label{fig:2_g}
%\end{figure}

\section{Acknowledgements}
We acknowledge the support of the European Research Council and the Spanish Ministry of Economy and Innovation through the JAE-Predoc Program (J. C.), Grants No. FIS2011-23713 (P. S.-J), FIS2012-33521 (R. A.), FIS2013-47328 (E. P.) and the Ramón y Cajal Program (E. P).
\bibliography{biblio}
\clearpage
\appendix
%\include{Appendix/Appendix1}
%\section*{Supplementary Material}
%\label{supplement}

\section{The SNS junction model}
\label{appA0}
In this appendix we describe the model we use for SNS junctions. 
\subsection{Tight-binding discretisation}
For computation purposes, we consider a discretisation of the continuum model Eq.\,(\ref{Leq1}) for the Rashba nanowire into a tight-binding lattice with a small lattice spacing $a$.
Thus $H_{0}$ reads,
\begin{equation}
H_{0}\,=\,\sum_{i}c_{i}^{\dagger}\,h\,c_{i}\,+\,\sum_{<ij>}c_{i}^{\dagger}\,v\,c_{j}\,+\,\text{h.c}\,.\,
\end{equation}
where the symbol $<ij>$ means that $v$ couples nearest-neighbor $i,j$ sites. This discretisation transforms Eq.\,(\ref{Leq1}), in terms containing on-site energy $h$ and into nearest-neighbor hopping matrices $v$ which arise from the momentum operator $p$,
\begin{equation}
\label{hopp}
\begin{split}
h_{ii}&\,\equiv\,h\,=\,
\left[
\begin{array}{cc}
2t\,-\,\mu&B\\
B& 2t\,-\,\mu
\end{array}
\right]\,,\\
h_{i+1,i}&\,\equiv\,v\,=\,
\left[
\begin{array}{cc}
-t&\frac{\hbar}{2a}\alpha_{so}\\
-\frac{\hbar}{2a}\alpha_{so}& -t
\end{array}
\right]\,=\,h_{i,i+1}^{\dagger},
\end{split}
\end{equation}
are matrices in spin space and $t=\hbar^{2}/2m^{*}a^{2}$. 

\subsection{The SNS junction model}
The Hamiltonian of the full system considering the proximized NW regions as left and right superconducting leads (see discussion at the beginning of section \ref{ElevelsSNS}A) is given by
\begin{equation}
\label{long1}
h_{SNS}\,=\,\left[
\begin{array}{ccc}
h_{S_{L}}&h_{S_{L}N}&0\\
h_{S_{L}N}^{\dagger}&h_{N}&h_{N S_{R}}\\
0&h_{N S_{R}}^{\dagger}&h_{S_{R}}
\end{array}
\right]\,,
\end{equation} 
where $h_{S_{i}}$ is the Hamiltonian of the superconducting lead $i=L/R$ that we consider to be the same, $h_{S_{i}N}$ the Hamiltonian that couples the  superconducting lead $i$ to the normal region, while $h_{N S_{i}}$ the Hamiltonian that couples the normal region to the lead $i$. 
These coupling matrices are non-zero for adjacent sites that lie at the interfaces of the superconducting leads and of the normal region, only.
This coupling is parametrized by a hopping matrix $v_{0}=\tau v$ between the sites that define the interfaces of the SNS junction, where $\tau\in[0,1]$. %Experimentally, it is hard to reach perfect contact in SNS junctions. 
A tunnel junction can be modelled by considering $\tau<<1$, while a full transparent junction with $\tau=1$.
%, although reaching this regime is experimentally delicate.
All the elements in the diagonal of matrix Eq.\,(\ref{long1}) have the structure of $H_{0}$ given by Eq. (\ref{Leq1}) taking into account that the superconducting lead regions have a Fermi energy $\mu_{leads}$ \new{(or $\mu_\mathrm{leads}$ for the normal transport study)}, while this is $\mu_{\mathrm{nw}}$ for the normal region. It is important to point out here that the matrix of Eq.\,(\ref{long1}) is of finite size since we are dealing with a fine size system.

Effects of superconductivity are induced by the pairing potential $\Delta(x)=\Delta{\rm e}^{i \varphi}$, thus leading to the Nambu description where the new Hamiltonian reads,
\begin{equation}
\label{superlong}
H\,=\,\left[
\begin{array}{cc}
h_{SNS}&\Delta(x)\\
\Delta^{\dagger}(x)&-h_{SNS}^{*}
\end{array}
\right]\,.
\end{equation}
The superconducting pairing potential in the previous Hamiltonian equation, that corresponds to the full system, must have the same structure as the SNS Hamiltonian, $h_{SNS}$, thus
 \begin{equation}
\label{long2}
\begin{split}
\Delta(x)=&\left[
\begin{array}{ccc}
\Delta_{S_{L}}&0&0\\
0&\Delta_{N}&0\\
0&0&\Delta_{S_{R}}
\end{array}
\right]\\
=&\left[
\begin{array}{ccc}
\Delta_{0,S}{\rm e}^{{\rm i}\varphi_{L}}&0&0\\
0&0&0\\
0&0&\Delta_{0,S}{\rm e}^{{\rm i}\varphi_{R}}
\end{array}
\right],
\end{split}
\end{equation} 
where $\Delta_{N}=0$ since in the normal region the superconducting correlations are absent. 

Superconductivity is induced by an s-wave pairing potential $\Delta(x)$ that couples particles of different spin and momenta.
So that, $\Delta_{0,S}$ is given by 
\begin{equation}
\Delta_{0,S}\,=\,{\rm i}\sigma_{y}\Delta_{S}=\,{\rm i}\sigma_{y}\Delta\,.
\end{equation}
%We always refer \new{to $\Delta_{S}$ as $\Delta$}.
%So far we have explained the model we employ to study single channel  one dimensional SNS junctions. 
\subsection{Induced superconducting pairing}
\begin{figure}[!ht]
\centering
\includegraphics[width=.45\textwidth,height=.1\textwidth]{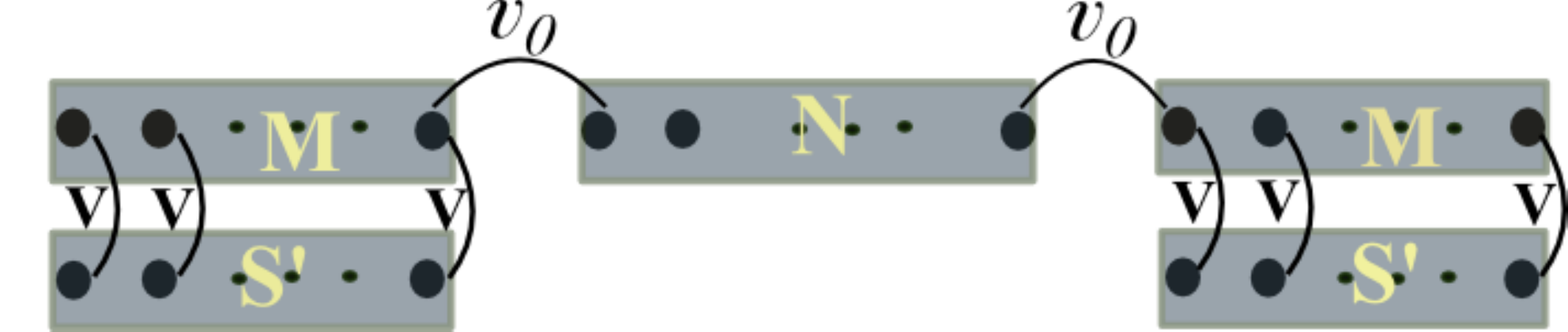} 
\caption{(Color online) A NW is divided in three normal regions (N) and (M), where the latter are coupled to a superconductor through $V$, while the coupling between (N) and (M) is controlled by $v_{0}$.}
\label{fig1App}
\end{figure}

A more realistic model consists on the following description. The full NW is divided in three sections: a central normal region (N) and two normal regions (M). See Fig.\,\ref{fig1App}. Each of the $M$ sections describe NW regions coupled to a superconductor which, to distinguish from the previous notation, we denote as $S'$. As opposed to the previous subsection, the full NW is now a normal system and the proximity effect comes now from the tunneling coupling between the superconductors and the $M$ normal parts of the NW.
In this case, the problem is described by the following Hamiltonian
\begin{equation}
\label{longsns2}
\hat{h}_{SNS}\,=\,\left[
\begin{array}{ccccc}
h_{S'_{L}}&h_{S'_{L}M}&0&0&0\\
h_{S'_{L}M}^{\dagger}&h_{M}&h_{MN}&0&0\\
0&h_{MN}^{\dagger}&h_{N}&h_{NM}&0\\
0&0&h_{NM}^{\dagger}&h_{M}&h_{MS'}\\
0&0&0&h_{M S'_{R}}^{\dagger}&h_{S'_{R}}\\
\end{array}
\right]\,,
\end{equation} 
where $h_{M}$ is a normal region of the same dimension as the superconducting one $S'$ and
$h_{S'_{i}M}$ is a diagonal matrix in site space that couples the superconductor $S'_{i}$ with the normal lead $M$. This coupling can be parametrized by the parameter $V$.
The superconducting pairing is then written in the same basis as $h_{SNS}$, 
\begin{equation}
\label{longpairing}
\begin{split}
\Delta(x)=&\left[
\begin{array}{ccccc}
\Delta_{S'_{L}}&0&0&0&0\\
0&0&0&0&0\\
0&0&0&0&0\\
0&0&0&0&0\\
0&0&0&0&\Delta_{S'_{R}}
\end{array}
\right]\,,
\end{split}
\end{equation} 
where $\Delta_{S'_{i}}=\Delta_{S'}{\rm e}^{{\rm i}\varphi_{i}}$ with $i=R,L$ describe the bulk s-wave superconducting leads. 

\begin{figure}[!ht]
  \label{locMaj}
\centering
\includegraphics[width=.3\textwidth]{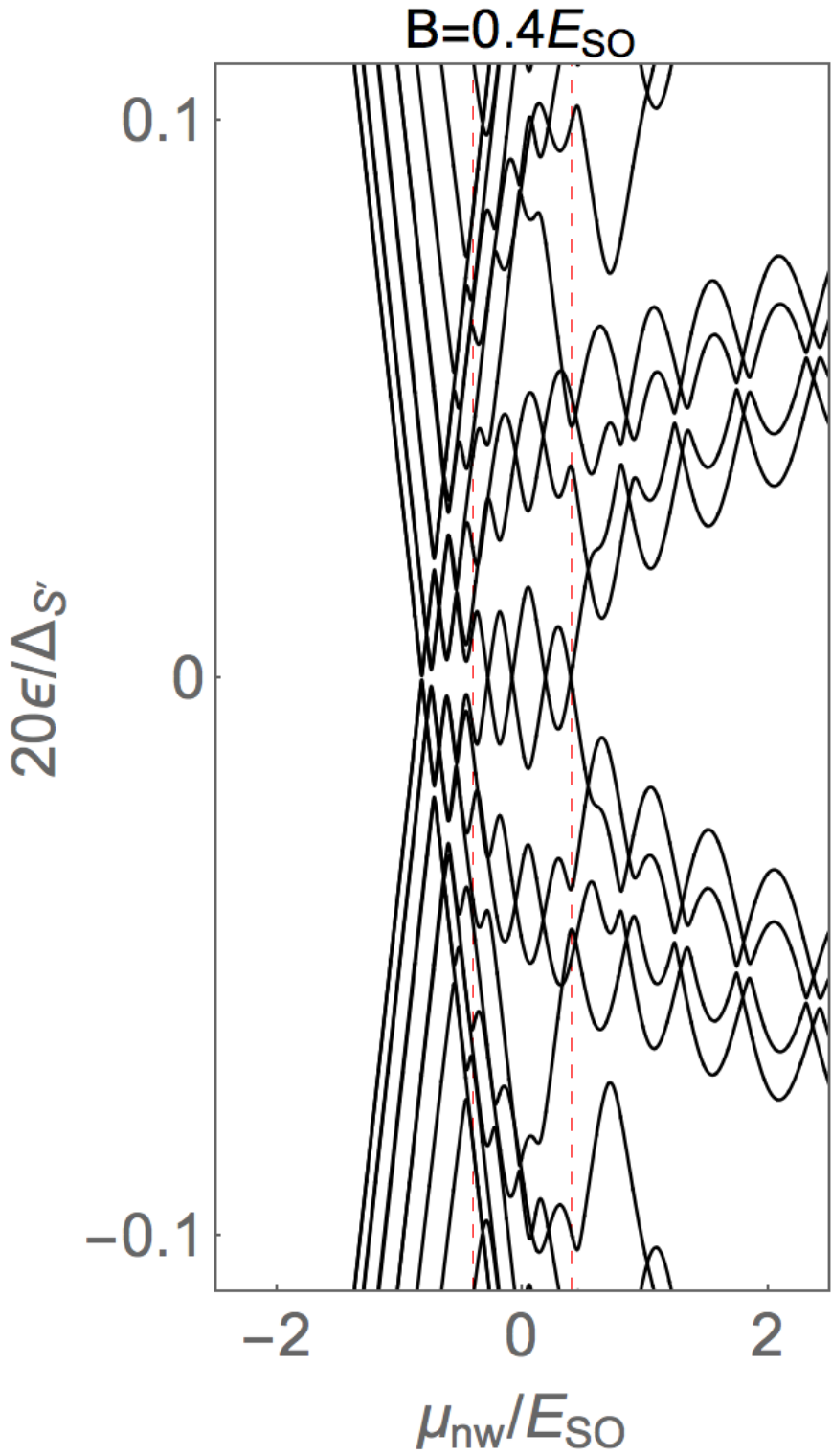} 
\caption{Energy levels at $\varphi=0$ as function of $\mu_{nw}$ for a long junction $L_{nw}=4\mu$m for a fixed Zeeman field. Parameters: $E_{SO}=0.05$meV, $\mu_{leads}=10E_{SO}$, $L_{S}=2\mu$m, $V=20E_{SO}$ and $\Delta_{S'}=20\Delta=5$meV. The rescaled $y$ axis explicitly shows that the relevant energy scale is not the original bulk gap included in the calculation $\Delta_{S'}$ but rather $\Delta$, in agreement with Fig. \ref{fig11}c.}
\label{fig4App}
\end{figure}

As described in the main text, the approximate description of the proximity effect in the previous subsection (Eq. \ref{long2}) is a good approximation provided that we are in a large gap limit and that the contact transparency is good. We have benchmarked the approximate solution of the previous subsection against the full proximity model in various relevant cases and always found good agreement in the correct parameter range. We here illustrate this point by showing a calculation using the full proximity effect model of Eq. \ref{longsns2} instead of the approximate model of Eq. \ref{long2}. In Fig.\,\ref{fig4App} we show results corresponding to the same physical situation we presented in Fig. \ref{fig11}c in the main text, the only difference being that the bulk gap in $S'$ is much larger than the induced gap used in the calculations of Fig. \ref{fig11}c ($\Delta_{S'}=20\Delta$). The overall behaviour of the subgap states in Fig.\,\ref{fig4App} is the same as in Fig. \ref{fig11}c (including the loops in the helical region described in the main text), demonstrating that the simplified model is indeed justified when the bulk gap is the largest energy scale. Importantly, note the rescaled $y$ axis which explicitly shows that the relevant energy scale is not the original bulk gap included in the calculation but the smaller value $\Delta=\Delta_{S'}/20$, in agreement with our previous claim.

\section{Model for the conductance}
\label{tightc}
In this part we make use of an effective model to describe the physics of Fano resonances. 
An effective \new{spinless} model based on Green's functions is constructed where two semi-infinite tight-binding chains (leads) are coupled through $V$ to a central region $\varepsilon_{d}$, formed by one site, that is additionally weakly coupled through $\tau<<V$ to a resonant level $
\varepsilon_{0}=\varepsilon_{d}-\varepsilon_{r}$, being $\varepsilon_{r}$ a fixed parameter that represents the separation between the quantum dot level and the resonant level (in principle this parameter mimics the role of the Zeeman splitting in our numerics). Consider that $a$ is the lattice constant and $t$ the hopping between sites in the leads.  The normal transmission, $T_{N}$, through a central system formed by one site can be calculated by using the Caroli's formula,
\begin{equation}
\label{transmq}
T_{N}(\omega)\,=\,4\,{\rm Tr}[\Gamma_{L}\,G^{r}\,\Gamma_{R}\,G^{a}]\,,
\end{equation}
where $G^{r(a)}$ is the retarded full system Green's function, and 
\begin{equation}
  \Gamma_{L(R)}(\omega)=\frac{\Sigma_{L(R)}^{r}(\omega)-\Sigma_{L(R)}^{a}(\omega)}{2i}\,,
\end{equation}
takes into account the influence of the leads on the central system through the left(right) L(R) self-energies $\Sigma_{L/R}$.
The full system Green's function can be calculated by using the Dyson's relation,
 \begin{equation}
 \begin{split}
  G^{r}(\omega)&=g_{0}^{r}(\omega)+g_{0}^{r}(\omega)\,\Sigma^{r}(\omega)\,G^{r}(\omega)=(G^{a}(\omega))^{\dagger}\\
 \text{or}\quad G^{r}(\omega)&=\left[[g_{0}^{r}(\omega)]^{-1}-\Sigma^{r}(\omega)\right]^{-1}\,,
  \end{split}
 \end{equation}
where $g_{0}^{r}$ is the retarded Green's function of the isolated central region (this central region can for instance be a quantum dot) without the influence of the leads and without the influence of the resonant level. \new{It reads}
\begin{equation}
g_{0}^{r}(\omega)\,=\,\frac{1}{\omega\,-\,\varepsilon_{d}\,+\,i\eta}
\end{equation}
where $\varepsilon_{d}$ is the onsite energy of the central region. 
 \begin{figure*}[!ht]
 \begin{minipage}[b]{\linewidth}
\centering
\includegraphics[width=.6\textwidth,height=.4\textwidth]{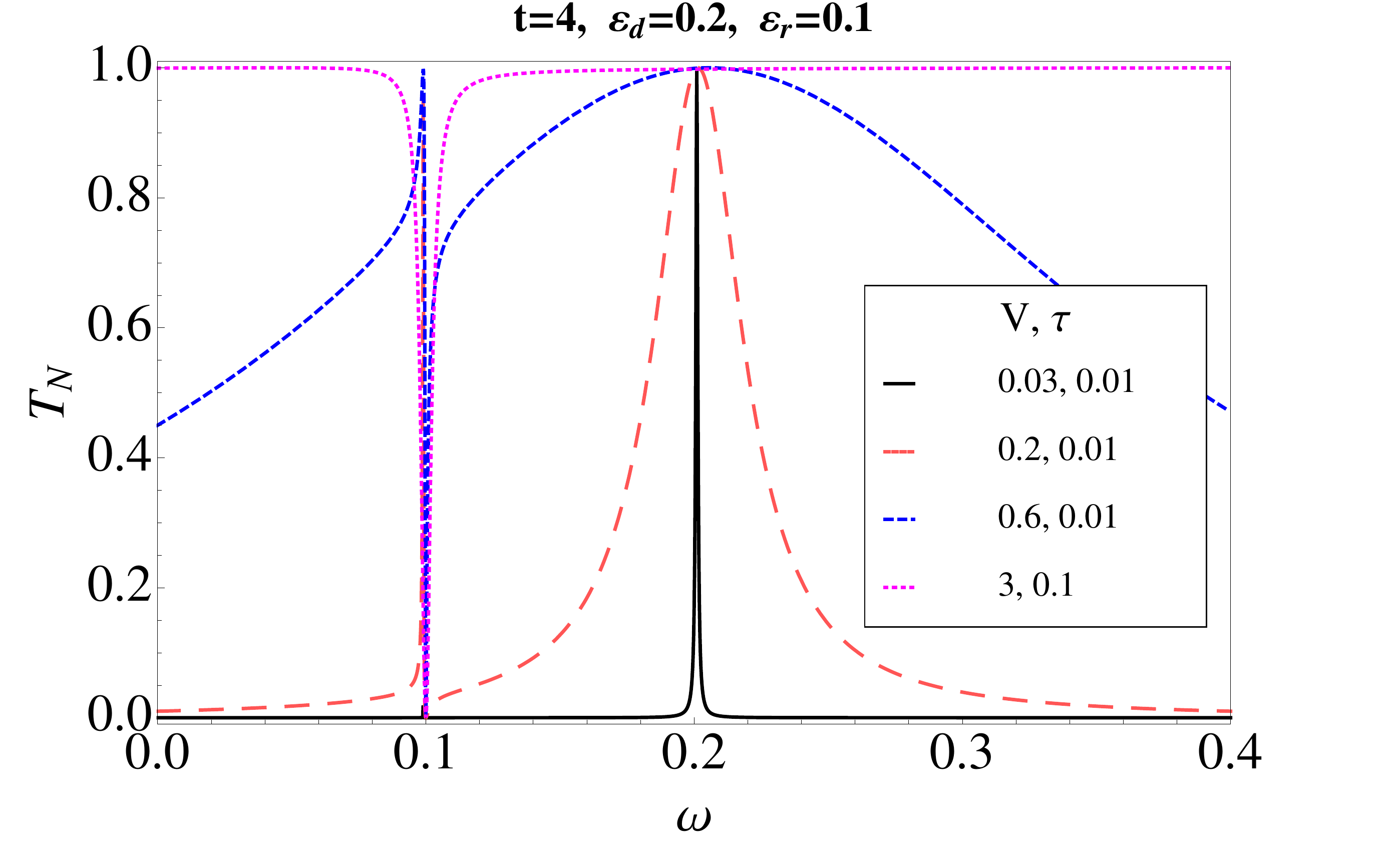} 
\caption{(Color online) Normal transmission for the system described in this subsection \ref{tightc}. Two tight binding semiinfinite chains (leads) coupled to central region formed by one site and where a resonant level is additionally weakly coupled to such central region. The hopping among sites in the leads is fixed and strong. By controlling the coupling to the leads $V$ and the one to the resonant level $\tau$ one observes that the normal transmission exhibit a resonant peak at the energy of the quantum dot for weakly coupling, however, by making the coupling to the leads stronger and leaving weak the one to the resonant level, $T_{N}$ develops a dip at the energy of the resonant level.
}\label{fig:TN}
\end{minipage}
\end{figure*}
 \begin{figure*}[!ht]
 \begin{minipage}[b]{\linewidth}
\centering
\includegraphics[width=.6\textwidth,height=.4\textwidth]{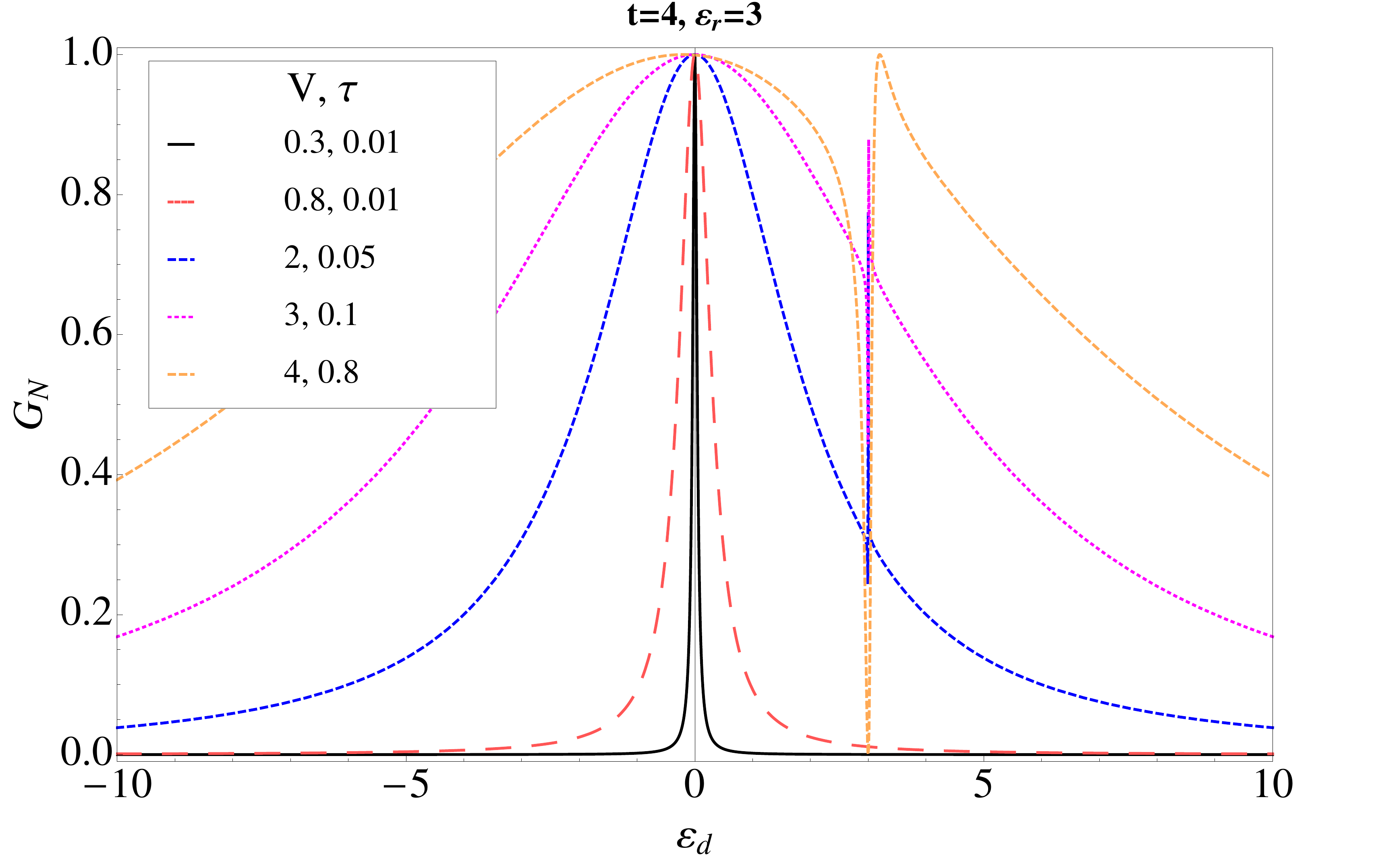} 
\caption{(Color online)Normal conductance across a central region attached to two semiinfinite tight-binding  chains (leads). In addition a resonant level is weakly coupled to such central region. The plots show the dependence of $G_{N}$ on the energy of the quantum dot $\varepsilon_{d}$. The hopping among sites in the leads is fixed and strong. By controlling the coupling to the leads and the one to the resonant level one observes that the normal conductance exhibit a resonant peak when $\varepsilon_{d}=0$, that is the Fermi energy of the leads $\omega_{F}=0$, for weakly couplings, however, by making the coupling to the leads stronger and leaving weak the one to the resonant level, $G_{N}$ develops a dip at the energy of the resonant level.
}\label{fig:GN}
\end{minipage}
\end{figure*}
The self-energy $\Sigma^{r}$,
\begin{equation}
\Sigma^{r}(\omega)=\Sigma_{L}^{r}(\omega)+\Sigma_{R}^{r}(\omega)+\Sigma_{res}^{r}(\omega)\,
\end{equation}
contain the effect of the left $\Sigma_{L}(\omega)$ and right $\Sigma_{R}(\omega)$ leads as well as the influence of the resonant level $\Sigma_{res}(\omega)$, respectively. Such self-energies are defined as follows,
\begin{equation}
\Sigma_{L(R)}^{r}(\omega)=t^{\dagger}\,g_{L(R)}^{r}(\omega)\,t
\end{equation}
where $g^{r}_{L(R)}$ is the retarded semi-infinite left (right) lead Green's functions. In principle such lead's Green's functions can be computed considering a recursive approach,
\begin{equation}
g_{L(R)}(\omega)\,=\,\frac{1}{\omega-h-t^{\dagger}\,g_{L(R)}(\omega)\,t}\,,
\end{equation}
$h=2t-\mu$ is the onsite energy in the leads. 
%, which in our calculation is set to zero. %(this is done in order to have an easy way to compare this calculations with the ones where p-wave superconductivity within the Kitaev framework in the topological phase where the onsite energy is set to zero). 
From previous equation one has,
\begin{equation}
|t|^{2}g_{L(R)}\,-\,(\omega-h)\,g_{L(R)}\,+\,1\,=0\,
\end{equation}
therefore,
\begin{equation}
g_{L(R)}(\omega)\,=\,\frac{1}{|t|}\left[\frac{\omega-h}{2|t|}\pm\sqrt{\left(\frac{\omega-h}{2|t|}\right)^{2}\,-\,1}\right]\,.
\end{equation}
Adding a convergence factor \new{to frequency, $\omega\rightarrow\omega\pm i\eta$,} one finds the retarded or advanced Green's function. \new{We have the following properties of $g_{L(R)}$},
\begin{widetext}
\begin{equation}
g_{L(R)}(\omega)\, = \begin{cases}
        \,\frac{1}{|t|}\left[\frac{\omega-h}{2|t|}-{\rm sgn}(\omega-h)\sqrt{\left(\frac{\omega-h}{2|t|}\right)^{2}\,-\,1}\right]\,,  &|(\omega-h)/2|t||>1\\
        \,\frac{1}{|t|}\left[\frac{\omega-h}{2|t|}\pm i\sqrt{1\,-\,\left(\frac{\omega-h}{2|t|}\right)^{2}}\right]\,,  & |(\omega-h)/2|t||<1
        \end{cases}
\end{equation}
\end{widetext}
where for the first case the density of states \new{$\rho_{0}=-\frac{1}{\pi}\mathrm{Im}g_{L(R)}$} is zero, while in the second case it exhibits a non zero value. These results allow us to obtain $\Sigma_{L}^{r}(\omega)$. The impurity self-energy $\Sigma_{res}^{r}$ reads,
\begin{equation}
\Sigma_{res}^{r}(\omega)\,=\,\frac{|\tau|^{2}}{\omega\,-\,\varepsilon_{0}\,+\,i\eta},
\end{equation}
where $\tau$ is the coupling of the resonant level to the system.
\new{With these expressions for the different self-energies, we may} compute $G^{r}$,
\begin{equation}
G^{r}(\omega)\,=\,\left\{[g_{0}^{r}(\omega)]^{-1}\,-\, \Sigma_{L}^{r}(\omega)\,-\,\Sigma_{R}^{r}(\omega)\,-\,\Sigma_{res}^{r}(\omega)\,\right\}^{-1}\,.
\end{equation}
The normal conductance $G_{N}$ is calculated from the transmission as,
\begin{equation}
G_{N}=\frac{e^2}{h}\int T_{N}(\omega)\left( -\frac{d\,f}{d\,\omega}\right)\, d\,\omega\,
\end{equation}
where by construction we have already in a spinless channel. 
Since we are interested in low temperature physics, $f(\omega)\approx\Theta(\omega_{F}-\omega)$, and $df/d\omega\approx -\delta(\omega_{F}-\omega)$. Therefore,
\begin{equation}
\begin{split}
G_{N}&=\frac{e^{2}}{h}\int T_{N}(\omega)\delta(\omega_{F}-\omega)\, d\,\omega\,\\
G_{N}&=\frac{e^{2}}{h}\, T_{N}(\omega_{F})\,,
\end{split}
\end{equation} 
where $\omega_{F}$ is the Fermi energy which \new{is the zero of energy} in our calculations.

%In computing the retarded and advanced Green's function, one has consider only the case where the local density of states is non zero. 
The aim of this part was to construct an effective model that contains the whole physics of our numerics where a resonance in the trivial phase and a dip in the helical phase the transmission develops.
Indeed, by plugging previous equations in the expression for the transmission and conductance, one ends up with the desired result that is plotted in Figs.\,\ref{fig:TN}, and \ref{fig:GN}.

In such plots, we consider a strong hopping $t$ between sites in the leads in comparison to the couplings $V$ and $\tau$. 
%Initially both the central region and the resonant level are weakly coupled to the leads and to the central region, respectively. 
\new{For weak coupling between leads and the central region a resonant tunnelling peak is obtained} at the energy of the central region $\omega=\varepsilon_{d}$. Upon increasing the coupling between the leads and the central region $V$ the resonant peak at $\varepsilon_{d}$ becomes broader and \new{a sharp Fano feature} emerges at the resonant impurity $\omega=\varepsilon_{r}$. \new{The new feature has the typical Fano structure of a zero followed by a peak, and arises from the interference of the two possible paths for the carriers, through the very broadened (strongly coupled) site at $\varepsilon_{d}$, and through the weakly coupled resonant level at $\varepsilon_{r}$. For strong enough coupling $V$, the $\varepsilon_{d}$ contributes with a uniform $e^2/h$ background to conductance, while the Fano feature becomes a pure dip to zero.}
%  $V$ the peak at $\varepsilon_{r}$ becomes more pronounced while the resonance at $\varepsilon_{d}$ even broader. 
%When the coupling to the leads is stronger, the peak at $\varepsilon_{r}$ changes to a kind of dip while the resonance at $\varepsilon_{d}$ tends to a huge broadening that disappears when $V$ is much higher. 

In conclusion, we have developed an effective model that contains the physics involved in our numerics where a resonance peak is present at the energy of the quantum dot for weakly coupled system. By increasing the coupling of the quantum dot to the leads a \new{Fano feature (dip to zero followed by a peak) appears in conductance} at the energy of the resonant level.  
%transforms into a dip when the coupling to the leads becomes strong.  
\section{Majorana localization length}
\label{Majorana-length}
\begin{figure}[!ht]
  \label{locMaj}
\centering
\includegraphics[width=.3\textwidth]{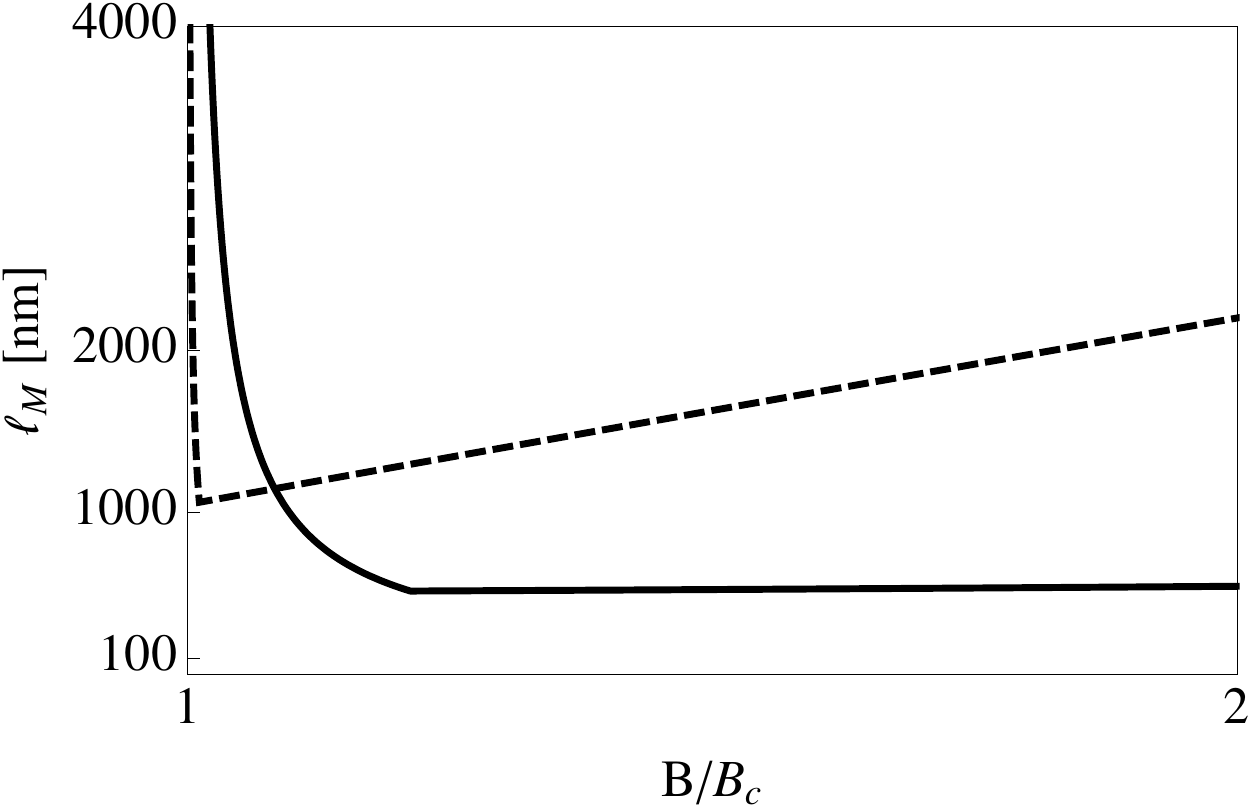} 
\caption{Majorana localization length $\ell_{M}$ as a function of the Zeeman field $B$ for $\alpha_{R}=\alpha_{0}$\, (dashed curve) and $\alpha_{R}=5\alpha_{0}$ (solid curve), where $\alpha_{0}=0.2\mathrm{eV \AA}$. They correspond, to spin-orbit lengths $l_{SO}\approx 200$nm and $l_{SO}\approx 40$nm, respectively. Rest of parameters $\mu=0.5$\,meV, and $\Delta=0.25$\,meV.}
\label{fig3App}
\end{figure}

The calculation of $\ell_{M}$ is carried out by solving the polynomial equation for the wave vector $k$ \, $k^{2}+4(\mu+C\alpha_{R}^{2})Ck^{2}+8\lambda C^{2}\Delta \alpha_{R} k+4C_{0}C^{2}=0$, where $C=m/\hbar^{2}$ and $C_{0}=\mu^{2}+\Delta^{2}-B^{2}$. Here, we point out that although the previous equation was derived  in Ref. \onlinecite{Lutchyn:PRL10} for a semiinfinite case, it gives reasonable values for the Majorana localization length. Indeed, in Fig.\,\ref{fig3App} one observes that $\ell_{M}$ linearly increases as one increases $B$ for realistic SOC (dashed line), while  it acquires smaller values and remains roughly constant for stronger SOC (solid curve). 
%Note that even for this latter case, which is somewhat unrealistic given the strong SOC, $\ell_{M}$ is considerably large at relevant Zeeman fields.

\end{document}